\documentclass{aa}

\usepackage{txfonts}
\usepackage{graphicx}
\usepackage{natbib}
\usepackage{subfigure}

\newcommand \cxc {{\it Chandra}}
\newcommand \xmm {{\it XMM-Newton}}

\begin{document}

\title{Six large coronal X-ray flares observed with Chandra}

\author{Raanan Nordon \inst{1} \and Ehud Behar \inst{2}}

\institute{Department of Physics, Technion, Haifa 32000, Israel;
nordon@physics.technion.ac.il
      \and Department of Physics, Technion, Haifa 32000, Israel; behar@physics.technion.ac.il}

\date{Received / Accepted}

\abstract{} {A study of the six largest coronal X-ray flares in
the \cxc\ archive is presented. The flares were observed on II
Peg, OU And, Algol, HR 1099, TZ CrB and CC Eri, all with the High
Energy Transmission Grating spectrometer (HETG) and the ACIS
detectors.} {We reconstruct an Emission Measure Distribution
$EMD(T)$, using a spectral line analysis method, for flare and
quiescence states separately and compare the two. Subsequently,
elemental abundances are obtained from the $EMD$.} {We find similar
behaviour of the $EMD$ in all flares, namely a large high-$T$
component appears while the low-$T$ ($kT <$ 2~keV) plasma is
mostly unaffected, except for a small rise in the low-$T$ Emission
Measure. In five of the six flares we detect a First Ionization
Potential (FIP) effect in the flare abundances relative to
quiescence. This may contradict previous suggestions that flares
are the cause of an inverse FIP effect in highly active coronae.}
{} 

\titlerunning{Six Large X-ray Flares}
\authorrunning{Nordon \& Behar}

\maketitle

\section{Introduction}

\begin{table*}[!t]
\begin{minipage}{\columnwidth}
\renewcommand{\thefootnote}{\arabic{footnote}}
\caption{\label{tab:obs_table} Observations used in this work.}
\begin{tabular}{c c c c c c c c c}
\hline \hline

Obs. ID &  HD & Other Name & Exposure (ks) &Start Date & Type & Radius & Distance (pc)$^{H}$ & P$_{orb}$ (d) \\

\hline

1451   & 224085  & II Peg & 43.3  &1999-10-17 23:28:28 & K2IV+?              & 2.2$^{S}$       & 42.34 &6.7 \\
1892   & 223460  & OU And & 96.9  &2001-08-11 00:18:00 & Single G1IIIe       & 9$^F$           & 135   &    \\
604    & 19356   & Algol  & 52.4  &2000-04-01 02:20:34 & B8V+G8III$^{B}$     & 2.8/3.54$^{B}$  & 28.46 &2.86$^B$\\
62538  & 22468   & HR1099 & 95.9  &1999-09-14 22:53:10 & G5IV+K1IV$^{S}$     & 1.3/3.9$^{S}$   & 28.96 &2.84\\
15     & 146361  & TZ CrB & 84.8  &2000-06-18 13:41:55 & F6V+G0V$^{S}$       & 1.22/1.21$^{S}$ & 21.69 &1.14\\
6132   & 16157   & CC Eri & 30.95 &2004-10-01 01:46:49 & K7Ve/M4$^{S}$       & 0.7/?$^{S}$     & 11.51 &1.56\\
4513   & 16157   & CC Eri & 89.45 &2004-10-01 20:54:39 & ...                 & ...             & ...   &... \\

\hline
\end{tabular}
H - The HIPPARCOS catalog \citep{Hipparcos} \\
S - \citet{Strassmeier1993} \\
B - \citet{Budding2004} \\
F - \citet{Fekel1986}
\end{minipage}
\end{table*}

The ongoing multi-wavelength research on stellar coronae has shown
that coronal activity in cool stars is closely related to the
magnetic activity. While in the solar case the coronal structures
are spatially resolved, one has to rely on accurate spectroscopy
in analyzing the unresolved stellar corona. Although showing
general similarity to the sun, some stellar systems exhibit
coronal activity 3-4 orders of magnitude more energetic than in
the sun, especially during large flares. Whether this indicates a
scaled up version of the solar coronal activity or a differently
structured corona is not yet clear.

Another effect which is not well understood is the abundance
variations between the photospheres and coronae of the stars. In
the solar case this has been known as the First Ionization
Potential (FIP) effect \citep{Feldman1992}, where compared to
their photospheric abundances, elements with low FIP
(under$\sim$10~eV) are over-abundant relative to elements with
higher FIP. Many systems show this effect, but some active systems
show an opposite effect where it is the high FIP elements who are
over abundant. This was labeled the Inverse FIP (IFIP) effect
\citep{Brinkman2001}. \citet{Audard2003} found that highly active
RS CVn binaries show an IFIP effect while less active systems show
either no effect or a solar FIP effect. \citet{Telleschi2005}
found a related result in a sample of solar like stars, where
abundances change from IFIP to FIP with the age (and decreasing
activity) of the star. This has led to the suggestion that
activity affects coronal abundances, possibly by flares that
evaporate high-FIP material from the lower chromosphere to the
corona \citep{Brinkman2001} and by electric currents that suppress
the diffusion of low-FIP species to the corona
\citep{Telleschi2005}. Observationally, however, the association to
date of flares with FIP or IFIP effects is ambiguous.
\citet{Gudel1999} and \citet{Audard2001} found an increase in low
FIP abundances during flares on the RSCVn UX Ari and HR 1099,
respectively. In some cases the variations in abundances were not
FIP-related \citep{Osten2003, Gudel2004}. In other cases, no
abundance variations were found at all, e.g., \citet{Maggio2000}
and \citet{Franc2001}, though the last two works are based on low
spectral resolution data.

The X-ray band offers several advantages in the study of stellar
flares. The high temperatures in the flares, typically with
$kT~\geq 1$~keV, mean that most of the emission is in the X-ray
band and the major emission lines of the highly ionized elements
fall in the 1.8 to 40~\AA\ range. This provides measurements of
line fluxes from up to 10 Fe K- and L- shell ionization degrees
(Fe XVII to Fe XXVI), as well as from the two K-shell ions of the
other common elements. While previous instruments limited plasma
emission models to two- or three- thermal components, line
resolved spectra from {\it Chandra} enable the reconstruction of a
more accurate distribution of plasma temperatures, as well as
better measurements of electron densities and abundances.

The reconstruction of the emission measure distribution ($EMD$)
that describes the plasma thermal structure, and the measurements
of abundances are difficult, not only because the two are entangled,
but also due to the poor mathematical definition of the problem as
shown by \citet{Craig1976}. Small variations of observed line
fluxes result in large variations in the derived $EMD$. Almost any
variation of the $EMD$ can be accommodated to produce the same
spectra, up to measurements uncertainties, if done on small enough
temperature scales. Consequently, many authors avoid setting
confidence intervals on their fitted $EMD$, or use $EMD$
smoothing, which makes the comparison of $EMD$s produced by
different methods a difficult task. The correlation between $EMD$
parameters and deduced abundances is often neglected or ignored.

In this work, we scanned the public {\it Chandra} archive for
bright flares that enable good line flux measurements. Six such
flares were found on the systems: II Peg, OU And, Algol, HR 1099,
TZ CrB \&\ CC Eri. While these are not the only flares in the
archive, they represent the brightest and best resolved flares
available. Some of these observations have been analyzed before,
but not always with regard to the flare and not in a way that
allows a comparison with other analysis methods. We refer to these
previous works in section~\ref{sec:Prev_works}.

We apply the same analysis methods, employing the derivation of a
continuous EMD and relative abundances, for the six targets in
flaring and quiescence states. The method is based on the one used
in \citet{Nordon2006}. Emphasis is given to the ability to
compare the results in a statistically significant way in order to
seek real variations in the thermal structure and abundances
between flare and quiescent states. In particular, we want to
establish whether the large flares had a statistically significant
effect on the coronal abundances and whether it is FIP-related.

\section{Observations}

\subsection{The Sample}
The \cxc\ public archive was searched for observations of cool
stars (spectral types A to M) in any system configuration,
featuring strong, long-duration flares. The main selection
criterion was the requirement for enough photons in the flare to
allow a detailed spectral analysis. So, while there are other
observed flares not included in this work, this sample represents
the best flares, in terms of photon counts, observed with {\it
Chandra}. All grating observations were examined, however, the
selected sample happens to contain only the ACIS+HETG instrument
configuration, which is the most common configuration used for
coronal targets.

The details of the selected observations and targets are
summarized in table \ref{tab:obs_table}. The data for CC Eri are
composed of two observations separated by a short time gap. The
flare occurred at the end of the first observation and the last
part of the flare decay was cut-off. Due to low flux in the
quiescence state, we integrated both the period before the flare
from the first observation, and the entire second observation for
the spectral extraction of the quiescence state.

\subsection{Light Curves and Spectra}

\begin{figure*}[!t]
\begin{center}
  \subfigure{\includegraphics[scale=0.35]{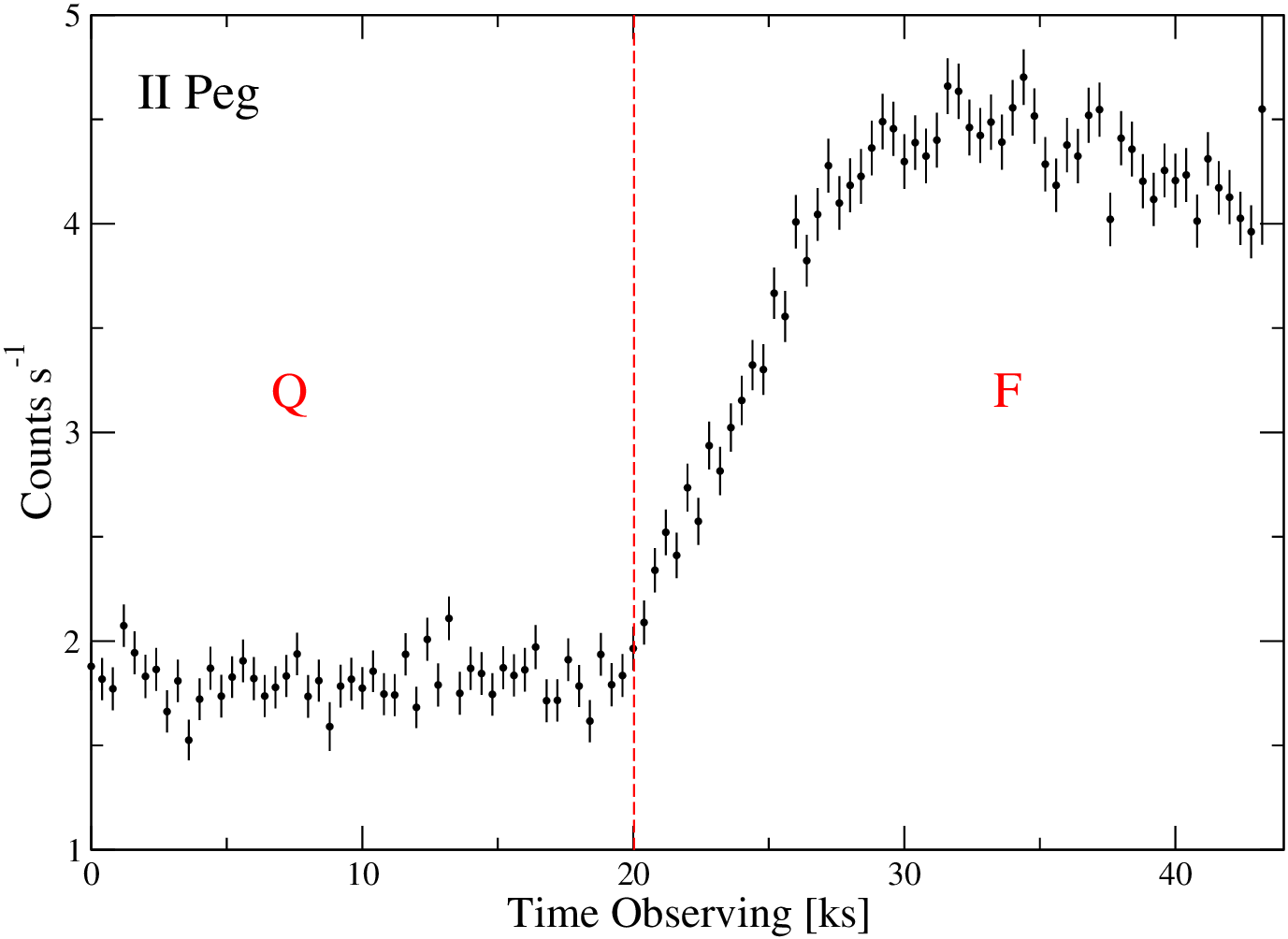}}
  \hspace{25pt}
  \subfigure{\includegraphics[scale=0.35]{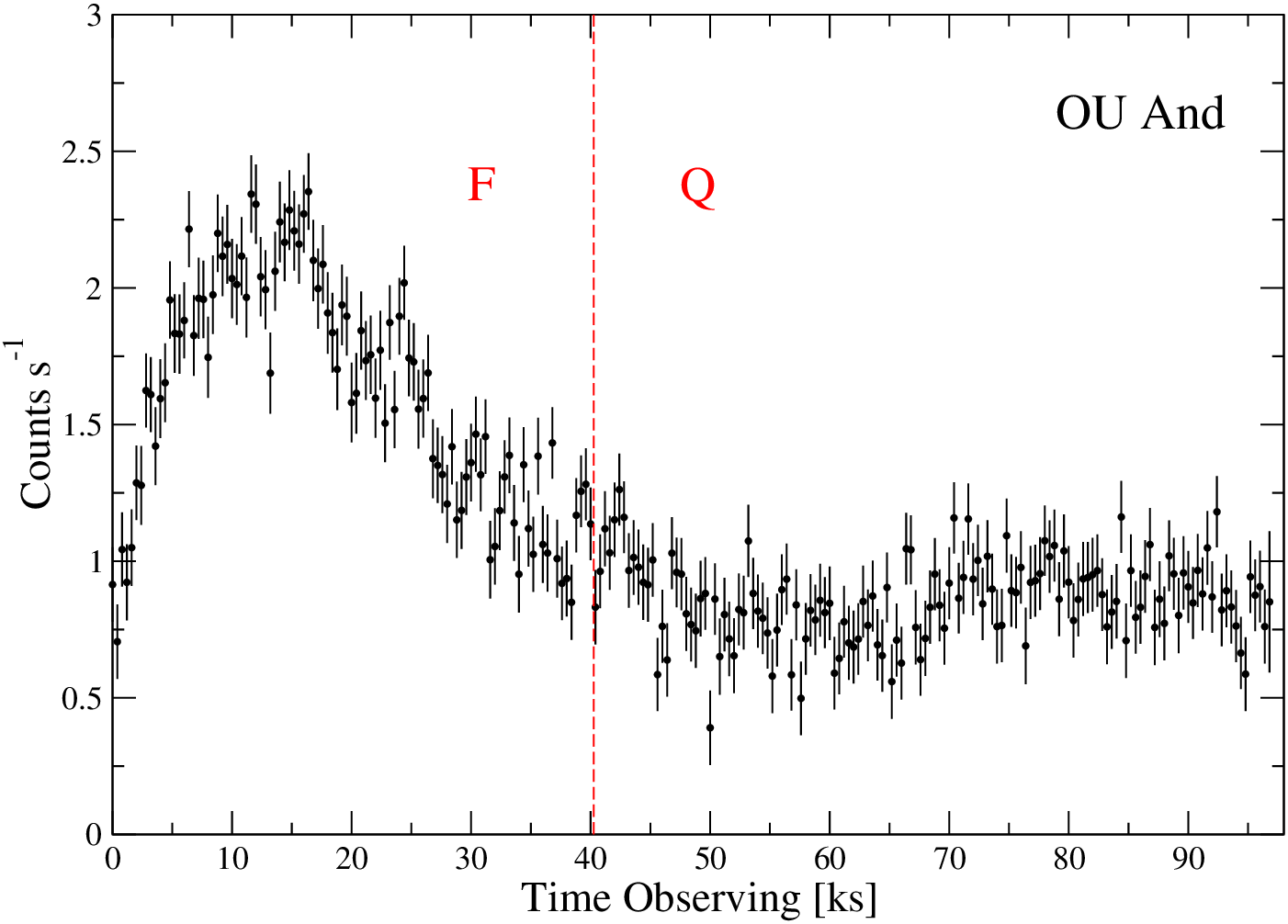}}
  \\
  \vspace{10pt}
  \subfigure{\includegraphics[scale=0.35]{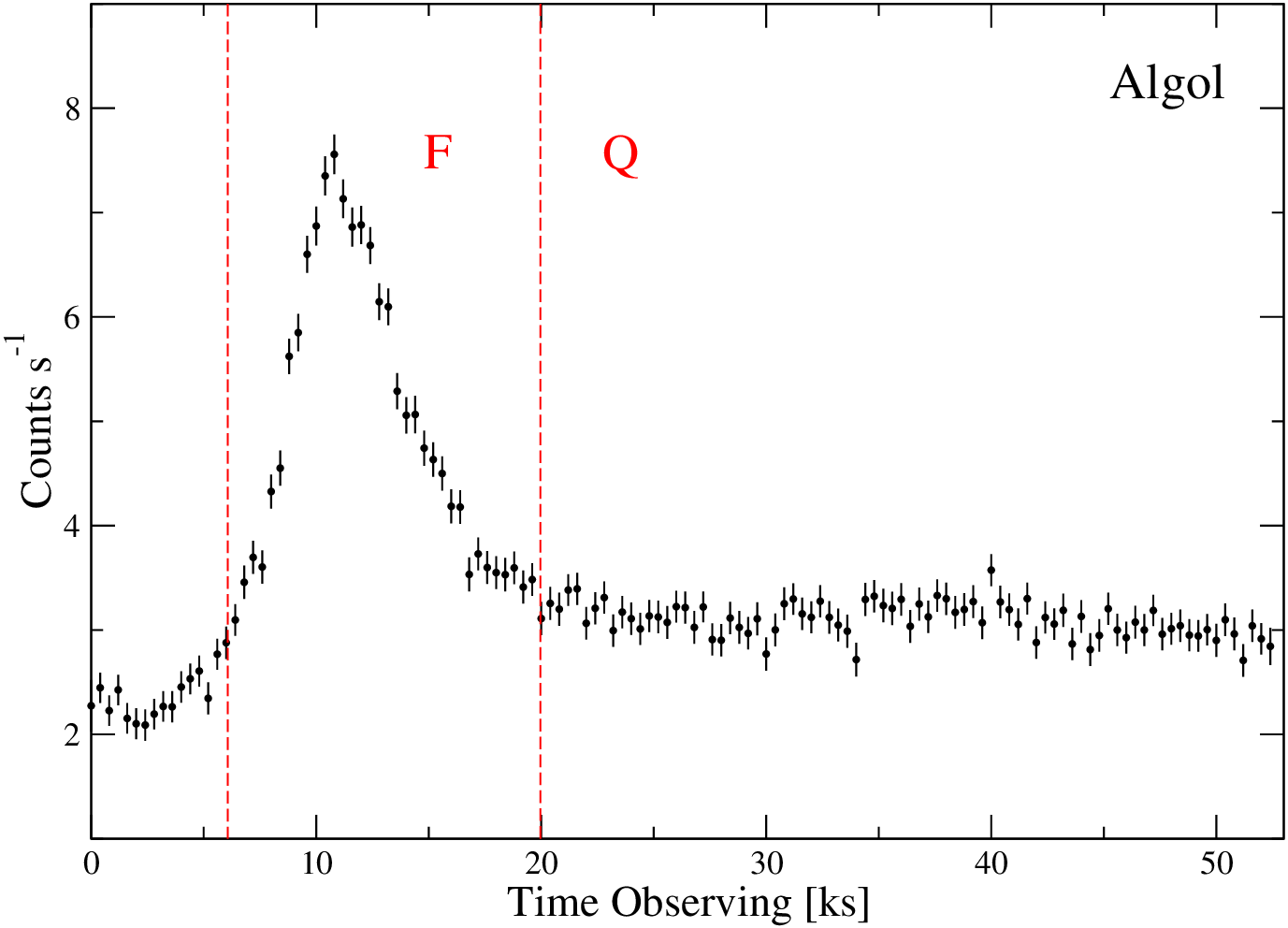}}
  \hspace{25pt}
  \subfigure{\includegraphics[scale=0.35]{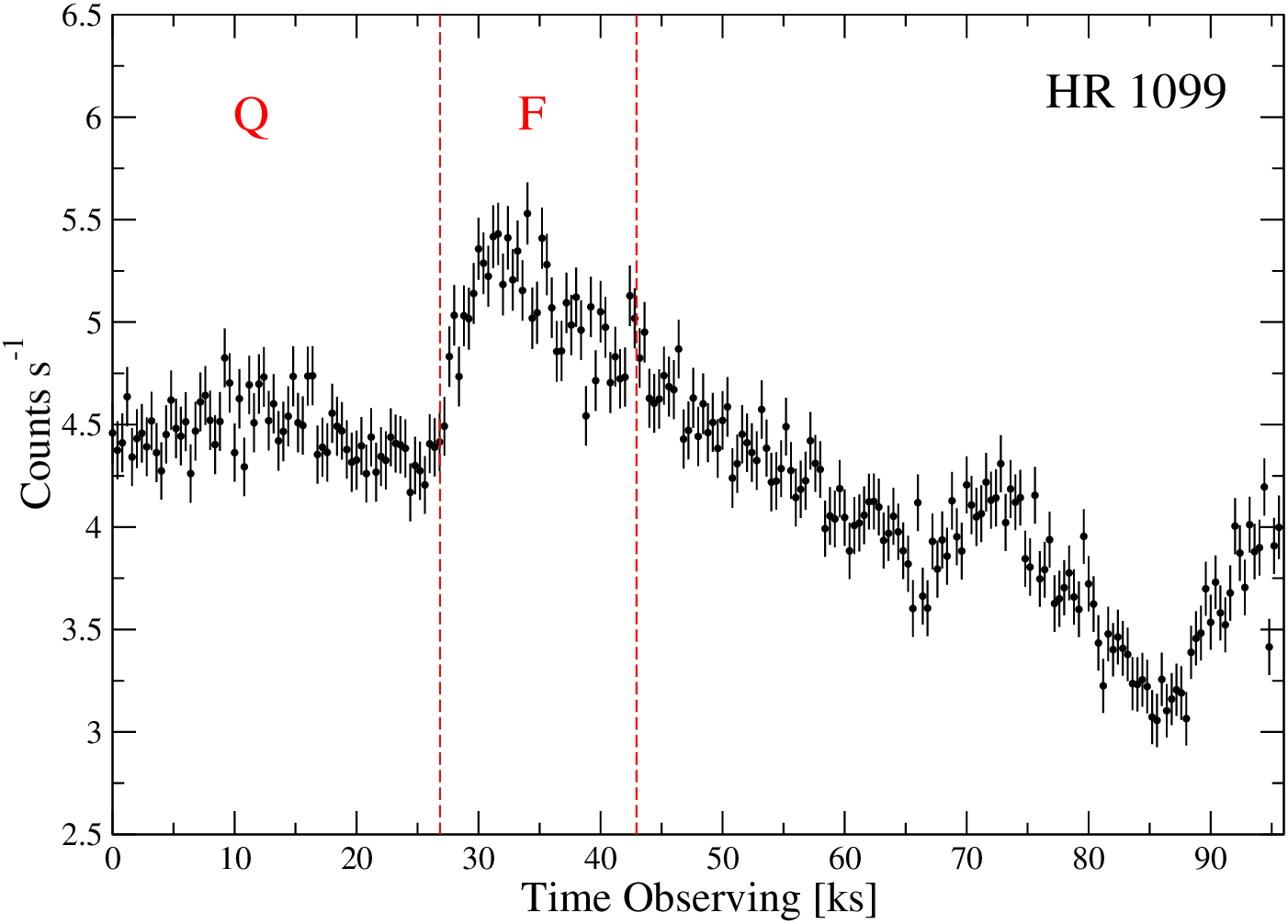}}
  \\
  \vspace{10pt}
  \subfigure{\includegraphics[scale=0.35]{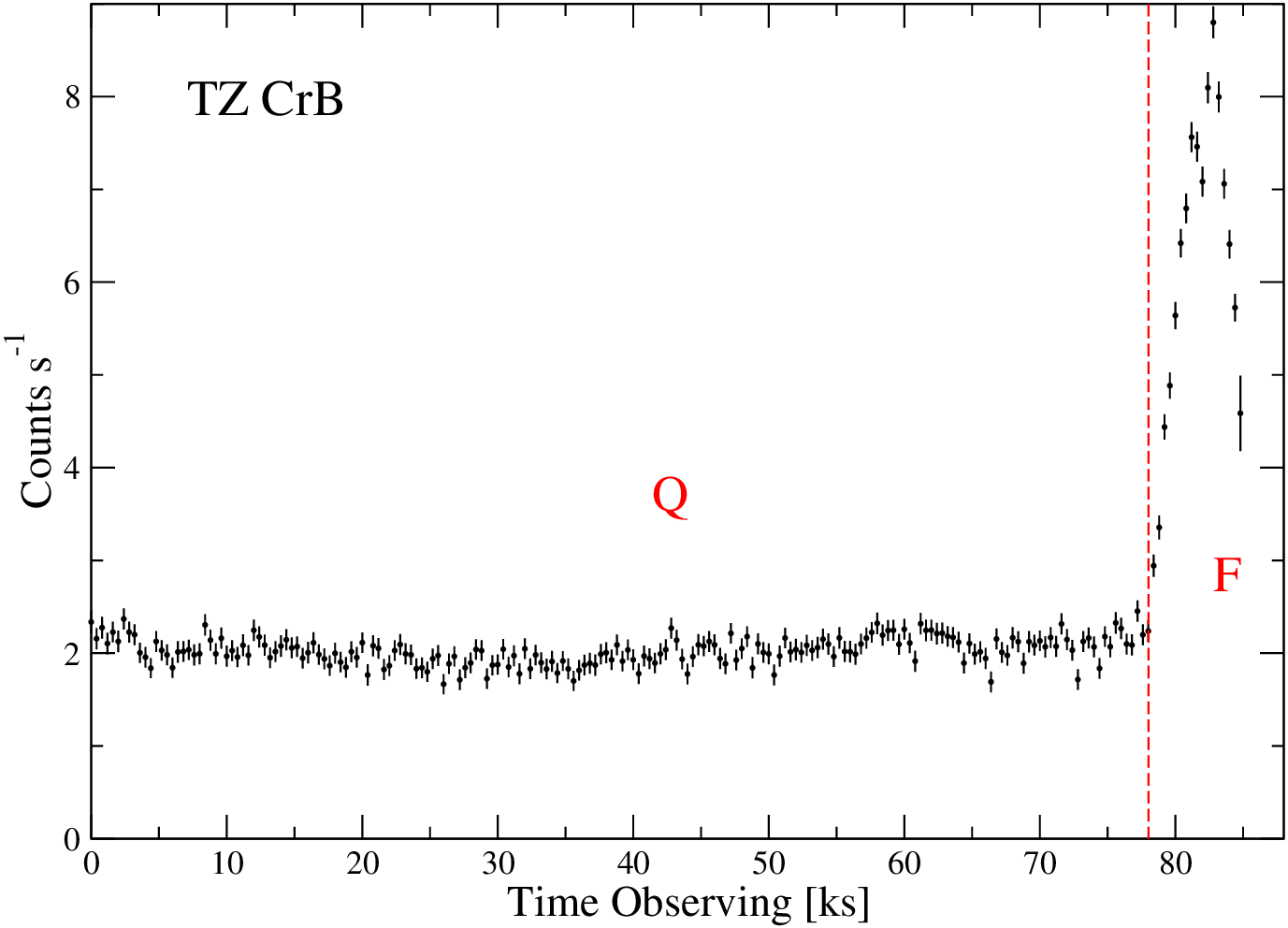}}
  \hspace{25pt}
  \subfigure{\includegraphics[scale=0.35]{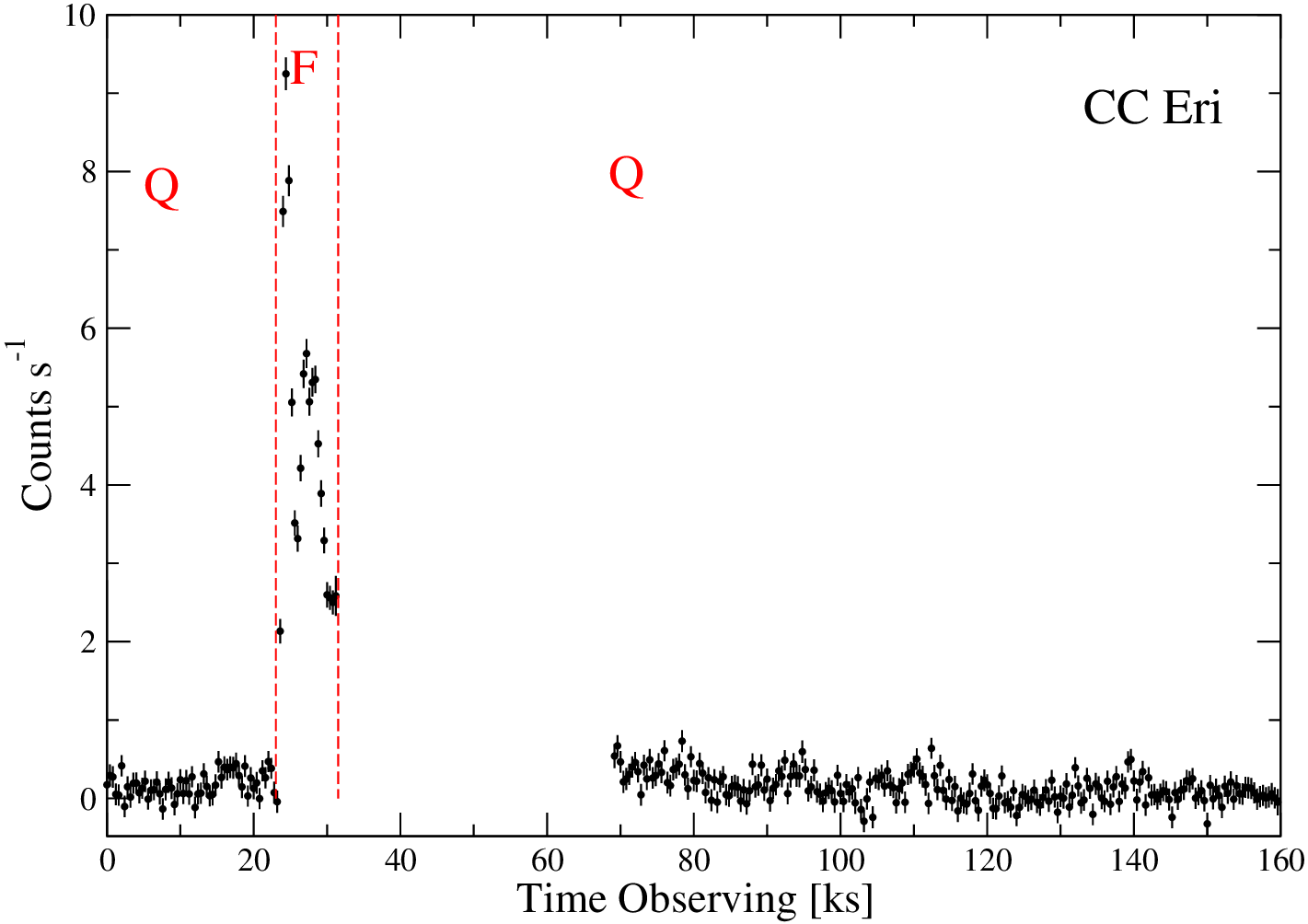}}
  \\
\caption{\label{fg:lc} Light curves of the observations using 400~s bins.
  All orders of diffraction are included and background is subtracted.
  The vertical dashed lines indicate the segments considered as flaring
  or quiescence and marked with letters F and Q respectively.
  The CC Eri observation
  consists of two nearly continuous observations.}
\vspace{20pt}
\end{center}
\end{figure*}

\begin{figure*}[!t]
\begin{center}
  \includegraphics[width=0.99\textwidth]{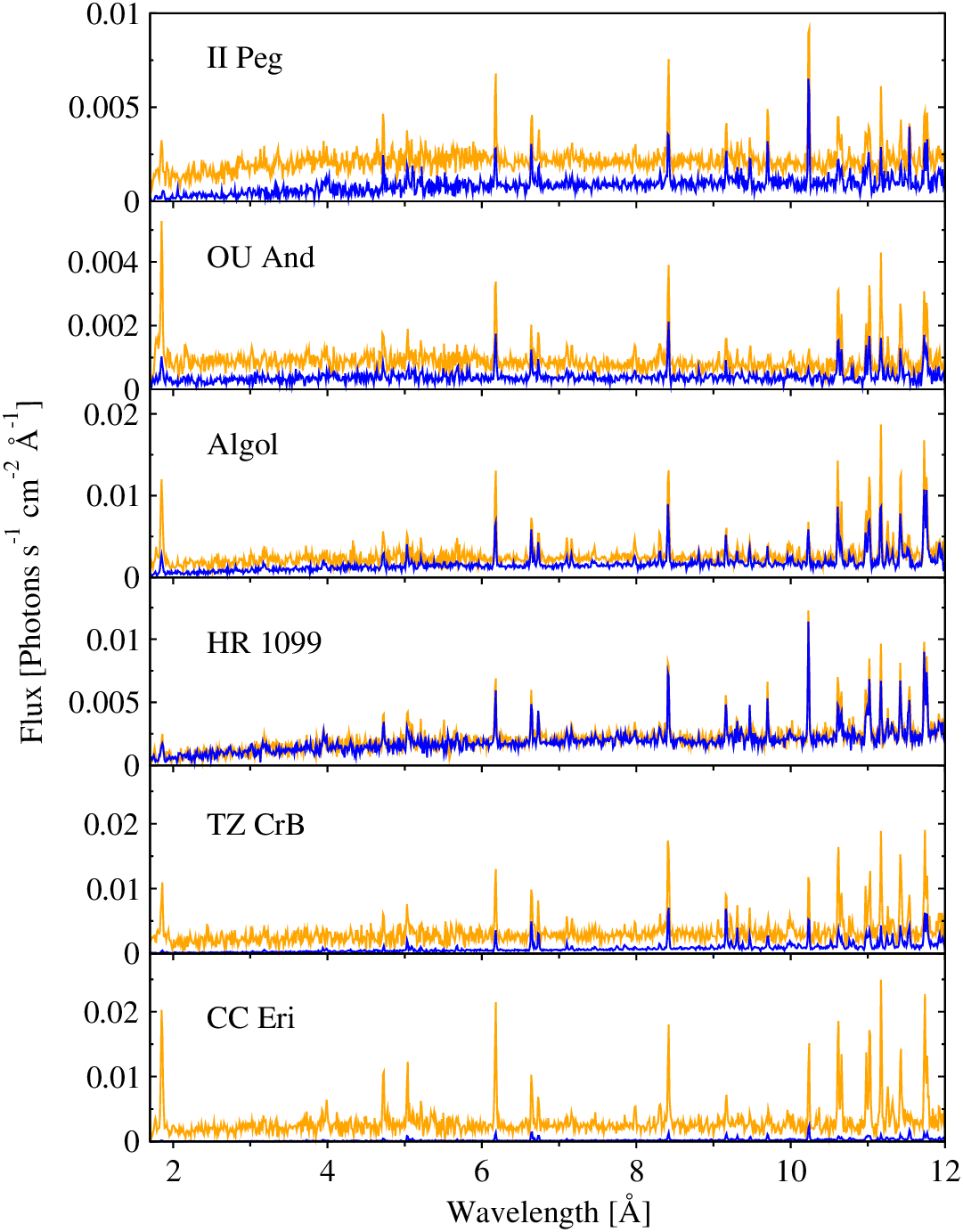}
\caption{\label{fg:Spectra1} Time averaged flare and quiescence spectra of the 6
targets in the 1.7-12~\AA\ range.
  HEG and MEG gratings 1st orders are combined at MEG resolution and 0.01~\AA\ bins. Orange is flare and blue is quiescence.}
\end{center}
\end{figure*}

\begin{figure*}[!t]
\begin{center}
  \includegraphics[width=0.99\textwidth]{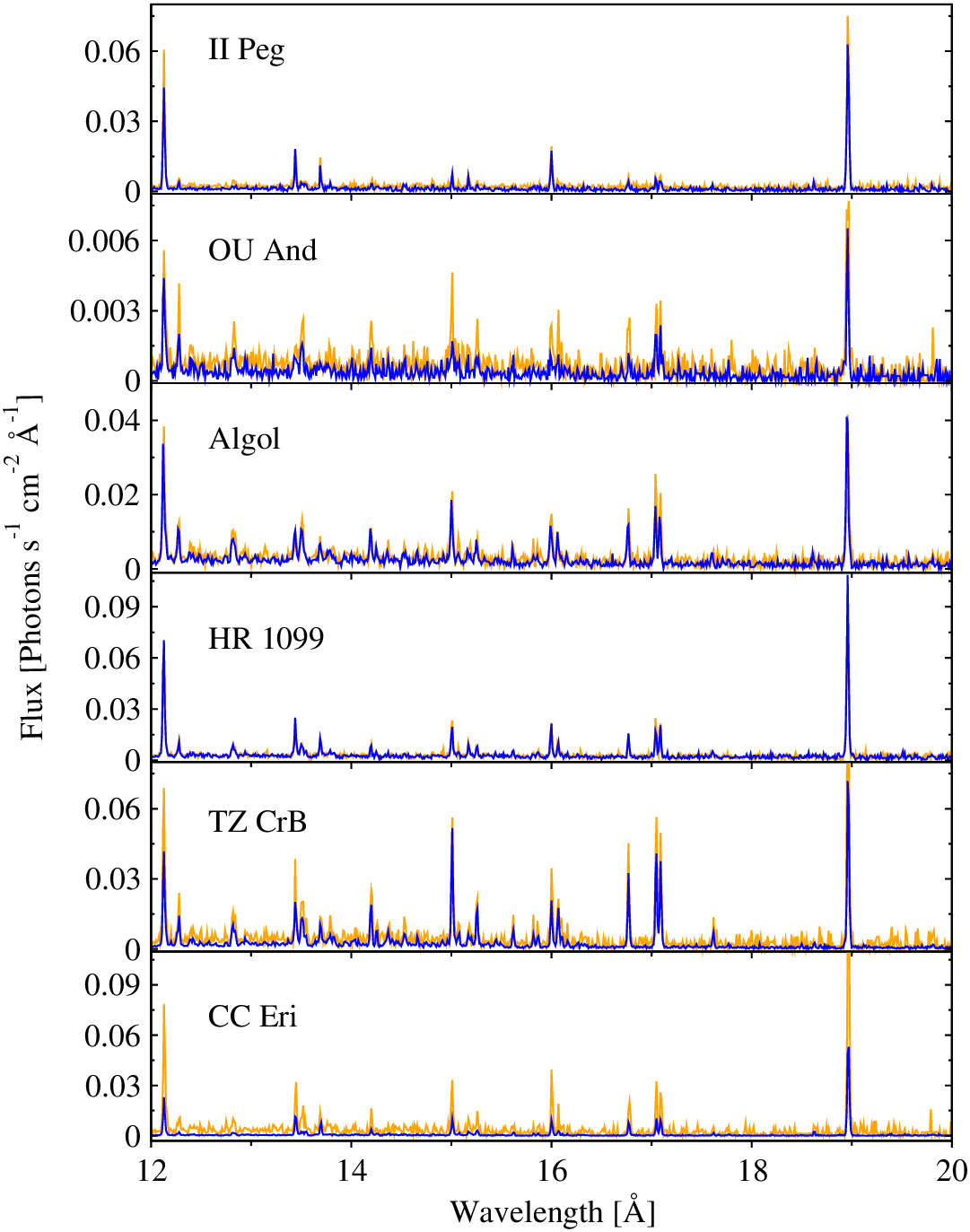}
\caption{\label{fg:Spectra2} Time averaged flare and quiescence spectra of the 6
targets in the 12-20~\AA\ range.
  HEG and MEG gratings 1st orders are combined at MEG resolution and 0.01~\AA\ bins. Orange is flare and blue is quiescence.}
\end{center}
\end{figure*}

Light curves were produced using combined counts of the High (HEG)
and Medium (MEG) energy grating arms of all orders as well
as the zero-order region. Background, though negligible, was
estimated and subtracted using off-source CCD regions. The light
curves are presented in Figure~\ref{fg:lc} using 400~s
bins. Segments used for flare and quiescence spectra extraction
are marked with {\it F} and {\it Q} respectively.

Flare and quiescent spectra are presented in
figures~\ref{fg:Spectra1}--\ref{fg:Spectra2} and correspond to the
time segments of the observations marked in
figure~\ref{fg:lc}. In the Algol observation, the time
segment before the flare was excluded as the system was still in
eclipse. The plots are in
0.01~\AA\ bins and use combined fluxed spectra of HEG and MEG.

\section{Modelling Method}
\subsection{Emission measure distribution}

While seeking the plasma parameters that reproduce the properly
measured fluxes of several selected lines,
we are particularly interested in examining the thermal structure
and elemental abundances independently. Therefore, we develop and
use a method that disentangles the mutual dependence in a simple
way. The continuum emission is problematic for two reasons; One
being the typical low level of continuum emission in quiescent
states. The other reason is the lack of unique features in the
continuum, 
which makes the distinction between thermal and
non-thermal components, or between high-$T$ components and low
metalicity, difficult and ambiguous. 
Better constraints are provided by the high energy end of 
the bremsstrahlung spectrum, but in cases of a wide temperature distribution
even the distinctive bremsstrahlung peak may be blurred. Moreover, in flares, 
the continuum turnover is often beyond the instrument band.
Thus, we avoid using the
continuum and rely on the more accurate line fluxes.

The observed line flux $F^{q}_{ji}$ of ion $q$ due to the atomic
transition $j \rightarrow i$ can be expressed by means of the
element abundance with respect to hydrogen $A_z$, the distance to
the object $d$, the line power $P^q_{ji}$, the ion fractional
abundance $f_q$ and the $EMD$ as:

\begin{equation}
  F^q_{ji} = \frac{A_z}{4 \pi d^2} \int_{0}^{\infty}{P^q_{ji}(T) f_q(T) EMD(T) \mathrm{d}T }
  \label{eq:lineflux}
\end{equation}

\noindent The line emissivity $\epsilon^q_{ji}(T) = P^q_{ji}(T)
f_q(T)$ describes ion and line specific parameters, while the
general plasma parameters are described by the $EMD$ as:
\begin{equation}
  EMD = n_e n_H dV/dT
  \label{eq:EMD_definition}
\end{equation}
\noindent Where $n_e$ and $n_H$ are the electron and hydrogen
number densities averaged over the volume of plasma in the
temperature interval [$T$, $T+\mathrm{d}T$].

The strongest unblended line from each ion is selected for the
analysis and density sensitive lines are avoided. Nonetheless, a
complete set of ionic spectra are used to account for residual
blending. The temperature dependence of the line emissivity
$\epsilon^q_{ji}(T)$ comes mainly from $f_q(T)$ and therefore a
second line from the same ion contributes little additional
information, except for more photon statistics. 
It could however, in some cases,  provide statistical compensation 
for uncertainties in the atomic parameters of the strong line.

We use the primary line from every Fe ion (Fe XVII to Fe XXV) in
the observed spectra to get a set of integral equations, 
whose solution yields the $EMD$ scaled by
the unknown Fe abundance:

\begin{equation}
  F^q_{ji} = \frac{A_{Fe}}{4 \pi d^2} \int_{T_0}^{T_{max}}{P^q_{ji}(T) f_q(T) EMD(T) \mathrm{d}T }
  \label{eq:lineflux_Fe}
\end{equation}

In order to include lines of other elements, but without the need
to fit their abundances, we use ratios of the He-like to H-like
line fluxes, from the same element. Thus, the element abundance
$A_z$ cancels out. This adds another set of equations that
constrain the shape of the $EMD$, but do not depend on the
abundances:

\begin{equation}
  R_z = \frac{ F^{He-like}_{ji} }{ F^{H-like}_{lk} }
  = \frac{ \int_{T_0}^{T_{max}}{P^{He}_{ji}(T) f_{He}(T) EMD(T) \mathrm{d}T } }
  { \int_{T_0}^{T_{max}}{P^H_{lk}(T) f_H(T) EMD(T) \mathrm{d}T }  }
  \label{eq:fluxratio}
\end{equation}

\noindent The X-ray spectra we use here include as many as ten Fe ions, but
no more than two ions from other elements. For every observation
we are able to use similar equations, but different specific ions,
depending on which lines have a sufficiently good signal.

The line fluxes and flux ratios are then fitted using the {\it
least squares best fit} method to solve this set of equations for the $EMD$, 
where the $EMD$ is expressed by a parameteric non-negative function of $T$
(see details in section~\ref{sec:EMD_param_and_fit}).
The minimization algorithms used are {\it Levenberg-Marquardt} and
{\it Nedler-Mead} (simplex) alternately and combined, taking the
best result. The fit yields the estimated shape of the $EMD$,
independent of any assumptions for the abundances, and is scaled
by the Fe abundance. In other words, we solve for the product of
$A_{Fe} \cdot EMD$. The integration in eqs.~(\ref{eq:lineflux_Fe},
\ref{eq:fluxratio}) is started at $kT_0=0.2$~keV and cut-off at
$kT_{max}=10$~keV, under and over which the $EMD$ is completely
degenerate, as no new lines emerge from these temperature regimes.
This means that some of the flux in high-$T$ lines may originate
from plasma at even higher temperatures, the effect of which will be
emission measure (EM) added to the last high-$T$ bin. 
The same applies to the lower temperature cut-off. Some emission 
in the low-$T$ lines may originate from temperatures lower than the cut-off.
For this reason, we avoid using lines with significant emissivity 
below the cut-off temperature, even if they are available (e.g. O VII). The coolest
forming ion we use is O VIII whose emissivity peaks at $\sim$0.25~keV and for
which over 95\% of the emissivity lies above the threshold.
\citet{Gu2006} showed that below 0.2~keV HETG
cannot constrain the $EMD$.
The abundance-independent approach, namely using abundance independent equations 
for fitting the $EMD$ and then using the $EMD$ to resolve the abundances, 
has been used in other works as well, 
for example: \citet{Schmitt2004} and \citet{Telleschi2005}.

The atomic data for the line powers are calculated using the
HULLAC code \citep{HULLAC}. The ionic abundances ($f_q$) for: Fe,
Ar, S, Si, Mg are taken from \citet{Gu2003}. \citet{Mazzotta1998}
is used for the other elements.

\subsection{Line fluxes}
In order to measure the line fluxes and solve for possible line
blending, we preform an ion-by-ion fitting to the spectra. The
line powers for each ion are calculated at its maximum emissivity
temperature and then passed through the instrument response. This
accounts for all of the lines and fixes the ratios of lines from a
given ion.

The observed spectrum is then fitted by sets of complete
individual-ion spectra simultaneously, together with a
bremsstrahlung continuum, composed of several discrete-temperature
components. This results in an excellent fit that accounts for all
the observed lines and blends. Since we ultimately use only
several selected lines for the analysis, we make sure that they
fit well and the fitting of the other lines serves for estimating
the blends in the selected lines (which is small to begin with).
The fitted continuum parameters are independent of the reconstructed $EMD$ and are
not used in the following analysis. The synthetic continuum serves only for
measuring line fluxes above it. Due to the high resolving power of the HETG that
produces narrow line profiles and the ionic spectra that include all lines 
that may contribute to a pseudo continuum, the 
local continuum level is reliable. The narrow line profiles of HETG also mean that
slight deviations in local continuum level have little effect on the measured line flux. 
This process is similar in principle to the one used in \citet{Behar2001} and in
\citet{Brinkman2001}.

The line fluxes measured from all the observations and used in the
following analysis are listed in tables~\ref{tab:line_fluxes1}-\ref{tab:line_fluxes3}
in the appendix.

\subsection{EMD parametrization and fitting}
\label{sec:EMD_param_and_fit}
Our goal is to compare the $EMD$ of the flare and quiescence
states. It is important to note that the solution for the $EMD$ is
not unique as is the case with integral equations of this sort
\citep{Craig1976}. On scales much smaller than the width of the
ions emissivity curves, or in temperature regions where the
relative emissivities vary slowly, there is no way of constraining
the shape of the $EMD$ from measurements alone. However, in order
to be able to compare different $EMD$ solutions, we need to have
meaningful confidence intervals. 
Currently, we do not have a
theoretical model that describes the $EMD$ of a full corona, although
ther are theoretical predictions for the $EMD$, under various assumptions.
We therefore make no assumptions as to the shape of the $EMD$ and 
choose to fit a {\it staircase} shaped function to allow
for local confidence intervals estimates. It also makes the
equations for the Fe line fluxes, which give the best constraints,
linear in the fitted parameters.

\begin{equation}
  EMD(T) = \{ C_n \hspace{10pt} T_n<T<T_{n+1} \: ; \: n=0...N-1 \}
\end{equation}

\noindent Where $N$ is the number of EM temperature bins, $T_n$
is the lower temperature of bin $n$. $C_n \geq 0$ is the parameter
to be fitted, which is the averaged $EMD$ over the bin. The
confidence intervals are calculated using the inverse $\chi^2$
distribution, meaning we search the parameter space for the
$\chi^2$ contour that gives a deviation from the best-fit that
corresponds to the requested confidence level. A $1 \sigma$
deviation is defined as the contour of $\chi^2_{min}+1$. The
selection of the number of bins and their widths is not trivial
and optimal $EMD$ binning can vary between different spectra. The
line emissivity curves have considerable widths and some extend to
temperatures much higher than their peak emissivity, resulting in
strong negative correlations between the $EMD$ in neigbouring
bins. Since, as discussed above, we are interested in meaningful
confidence intervals, we cannot use many narrow bins, as this will
result in excessive error bars. By making narrow bins they also
become closer in temperature, the emission contribution of
neighbouring bins becomes similar, thus increasing the degeneracy.
Ultimately, if meaningful confidence intervals are to be obtained,
the number of bins has to be kept small, and the tradeoff between
temperature resolution (number of bins) and constraints
(degeneracy of the solution), optimized. No smoothing algorithm is
applied to the $EMD$.

In addition, we fit an $EMD$ parameterized as an exponent of a
polynomial ensuring that the $EMD$ remains positive:
\begin{equation}
EMD(T) = \exp \left( \sum_{n=0}^{N}{C_n P_n(T)} \right)
\end{equation}
\noindent where $P_n(T)$ is an $n$-degree polynomial represented
as a Chebyshev polynomial for numerical convenience and summed up
to $N=7$. This is similar in principle to the method used by
\citet{Huenemoerder2001}, but with no smoothing. In this method,
local confidence intervals can not be produced, but we use it to
verify that our results do not depend on the $EMD$
parameterization.

\subsection{Integrated EM}
The physical measurable quantity is the observed line flux, which
results from the entire plasma (an integration over the $EMD$,
eq.~\ref{eq:lineflux}). Therefore, the meaningful quantity is the
integral of the $EMD$ over a range of temperatures. In other
words, the total EM in that range. Since emissivity curves are
smooth, uncertainties in the exact way the EM is distributed over
a temperature range, narrower than the emissivity widths, will not
have a significant effect on the total EM in that range. When
integrating and taking correlations into account, uncertainties
caused by the strong correlation between neighboring $EMD$ bins
that were integrated over, disappear. This results in much smaller
errors.

Temperature regions with little emissivity variations lead to poor
localization of the $EMD$ that in turn create spikes in the fitted
$EMD$ if no smoothing algorithm is used. Such spikes in the
solution are likely to appear when too narrow bins are used. This
reflects the fundamental mathematical instability of the solution.
The progressively Integrated EM as a function of
temperature (IEM($T$)) provides natural smoothing to these spikes as the
confidence intervals always vary smoothly. When comparing two
different $EMD$s (e.g., flare and quiescence states), since we
propagate the confidence intervals, the IEM indicates with greater
certainty where the two solutions differ. It also allows the
comparison of results obtained by different methods or by
different binning as it clears out correlation related
uncertainties. Over-binning, thus, which can make the $EMD$
meaningless, will not render the IEM unusable.

For the staircase $EMD$, the IEM is tracked at each stage of the
search for the parameters confidence intervals. This way, we get
good estimates for the upper and lower limits of the IEM curve,
inside a $\chi^2$ surface that represents a $1\sigma$ deviation.
We also integrate over the $EMD$ from the exponential model to
verify that the IEM is consistent regardless of the chosen $EMD$
parameterization.

\subsection{Abundances}
\label{sec:calculating_abundances} In order to extract the X/Fe
abundance ratios, we calculate the non-Fe line fluxes predicted by
the fitted, Fe-scaled, $EMD$.
\begin{equation}
  F^q_{ji} = \frac{A_{Fe}}{4 \pi d^2} \left( \frac{A_Z}{A_{Fe}} \right) \int_{T_0}^{T_{max}}{P^q_{ji}(T)
  f_q(T) EMD(T) \mathrm{d}T }
 \label{eq:calc_abund}
\end{equation}

\noindent The ratio of the actual measured fluxes to these
calculated fluxes gives the X/Fe abundance values. Note that for
calculating abundances, one line from each element is sufficient,
while for the $EMD$ equations we require two lines of different
ionization degrees. This is why we can estimate abundances for
elements not used in the $EMD$ fitting. 
Such is the case with Oxygen where we exclude O VII as it forms 
mostly at temperatures below our low-T cutt-off, but use O VIII
for the abundace.

Abundance uncertainties are directly related to the specific
element line flux errors, but they are also indirectly a result of
$EMD$ uncertainties. We take into account the latter by tracking
the upper and lower limits on the abundances as we fit the
$EMD$ and calculate its confidence intervals. This gives the
possible abundance values within a one $\sigma$ variation of the
$EMD$. Since $EMD$ is determined mostly by Fe and a combination of
other elements, it is justified to treat the two error
contributions as uncorrelated. We have also calculated the
abundances using the exponential $EMD$ model and verify that the
results of the two models are in good agreement.

We do not calculate the absolute abundances, i.e., the abundances
relative to hydrogen. The absolute abundaces are sensitive to the continuum, 
which is produced mainly by the less constrained high-T $EMD$. 
This is demonstrated in the work of \citet{Gu2006}
where different instruments observing the same source produce different 
absolute abundances, but similar relative abundances.
Since we are interested in abundance {\it variations}
during the flare, we use the relative (to Fe) abundances, obtained strictly from
line fluxes.

\section{Results}

\subsection{EMD and integrated EM}

\begin{figure*}[!t]
  \begin{center}
    \subfigure{\includegraphics[scale=0.35]{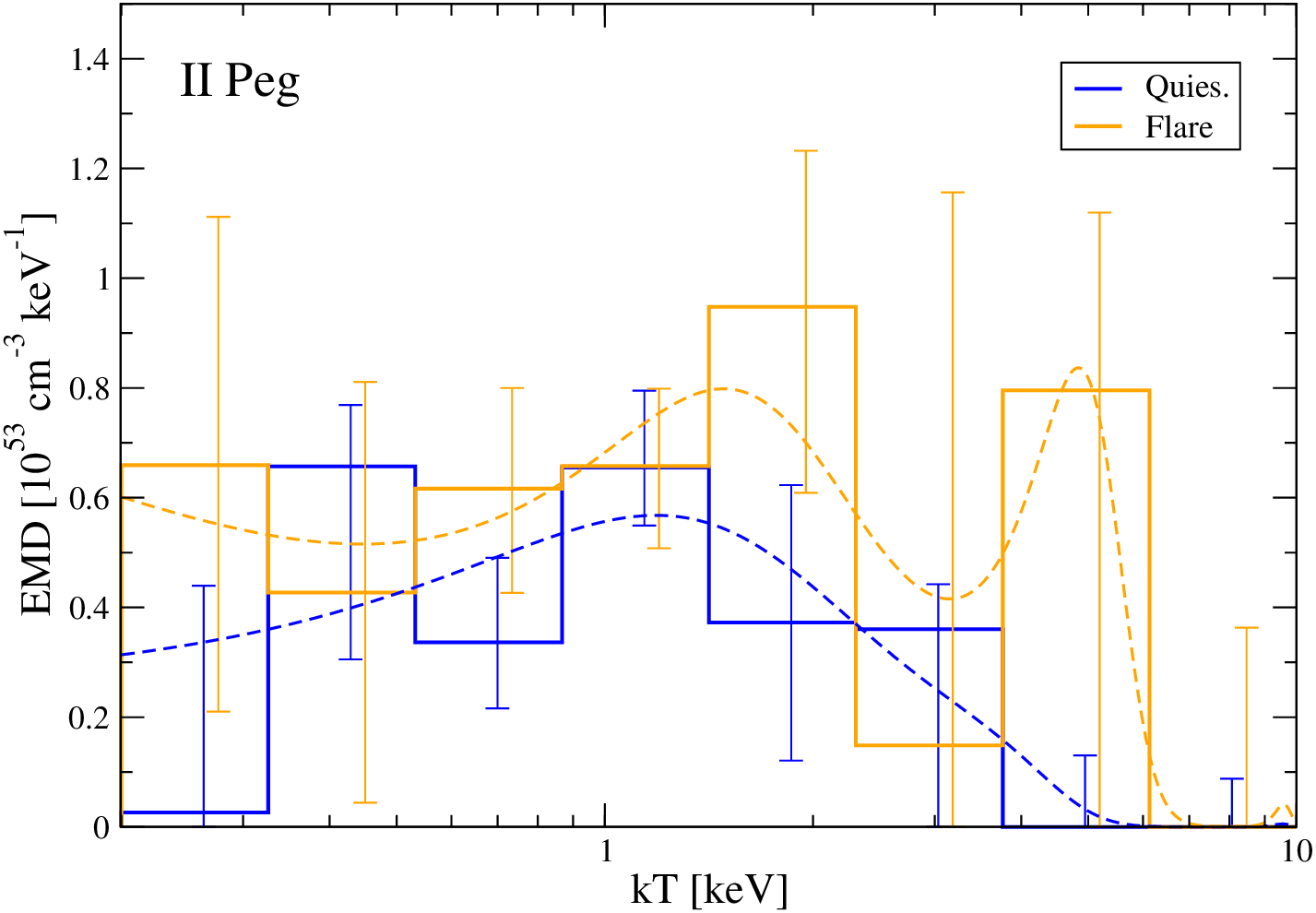}}
    \hspace{25pt}
    \subfigure{\includegraphics[scale=0.35]{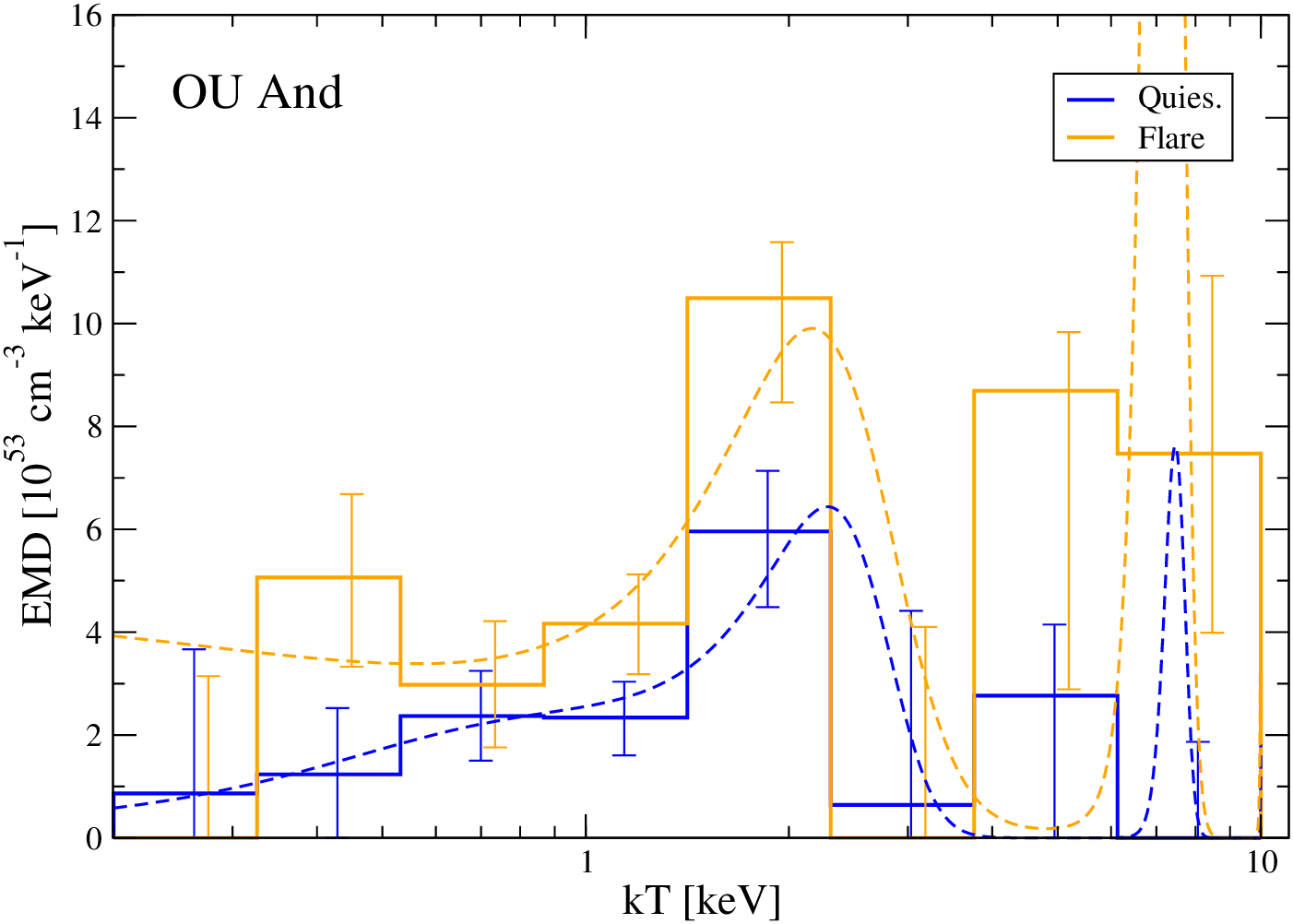}}
    \\
    \vspace{10pt}
    \subfigure{\includegraphics[scale=0.35]{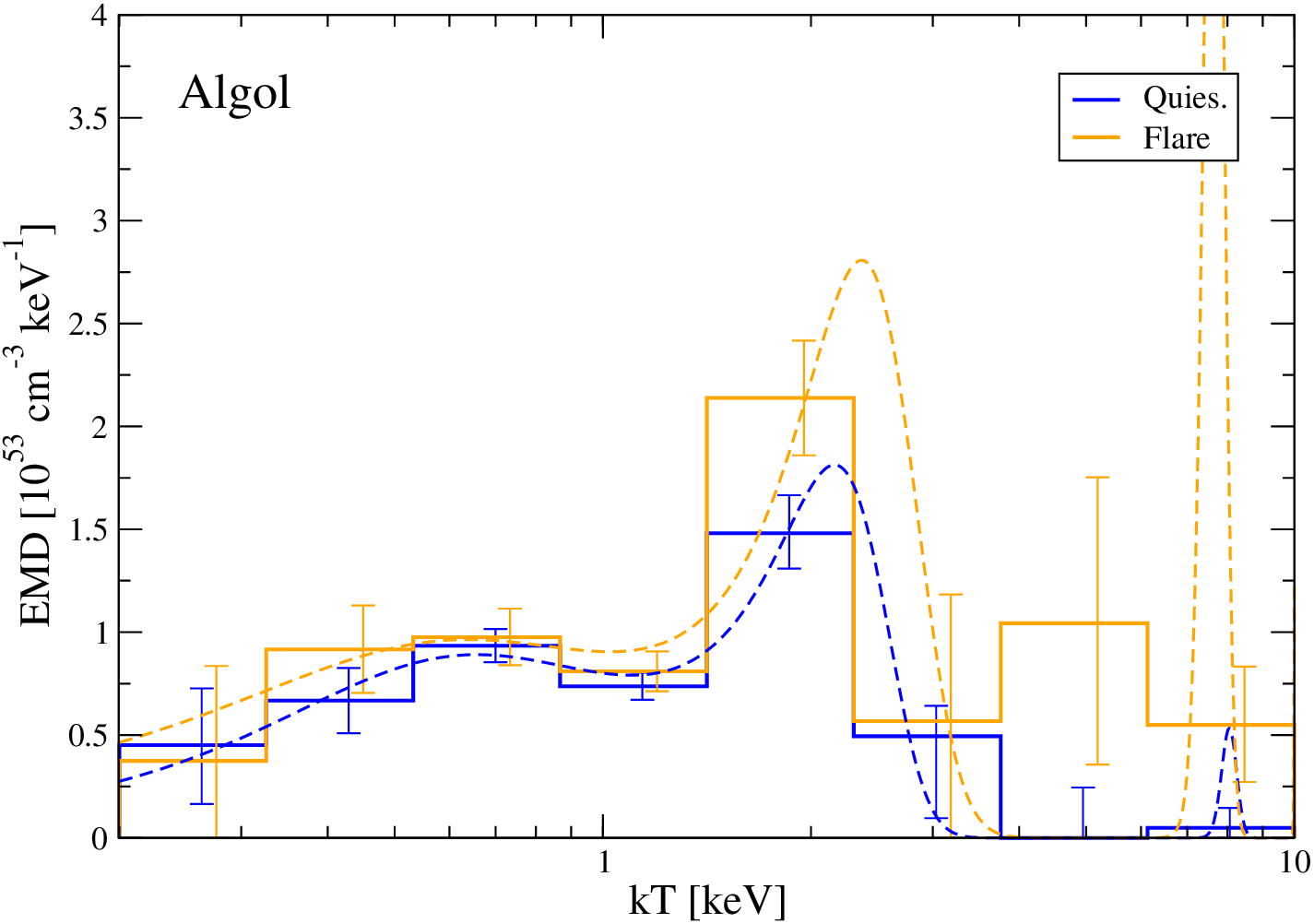}}
    \hspace{25pt}
    \subfigure{\includegraphics[scale=0.35]{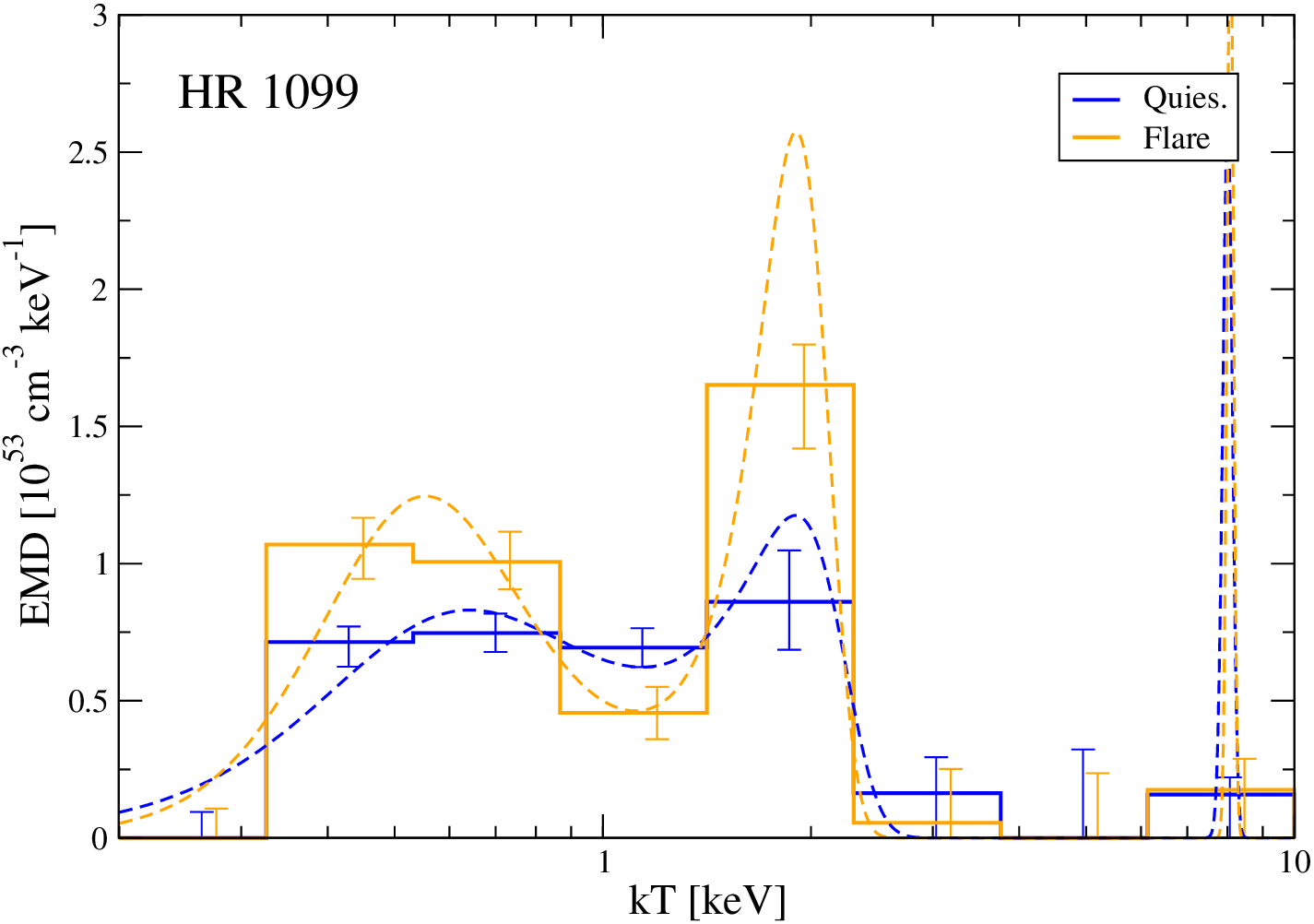}}
    \\
    \vspace{10pt}
    \subfigure{\includegraphics[scale=0.35]{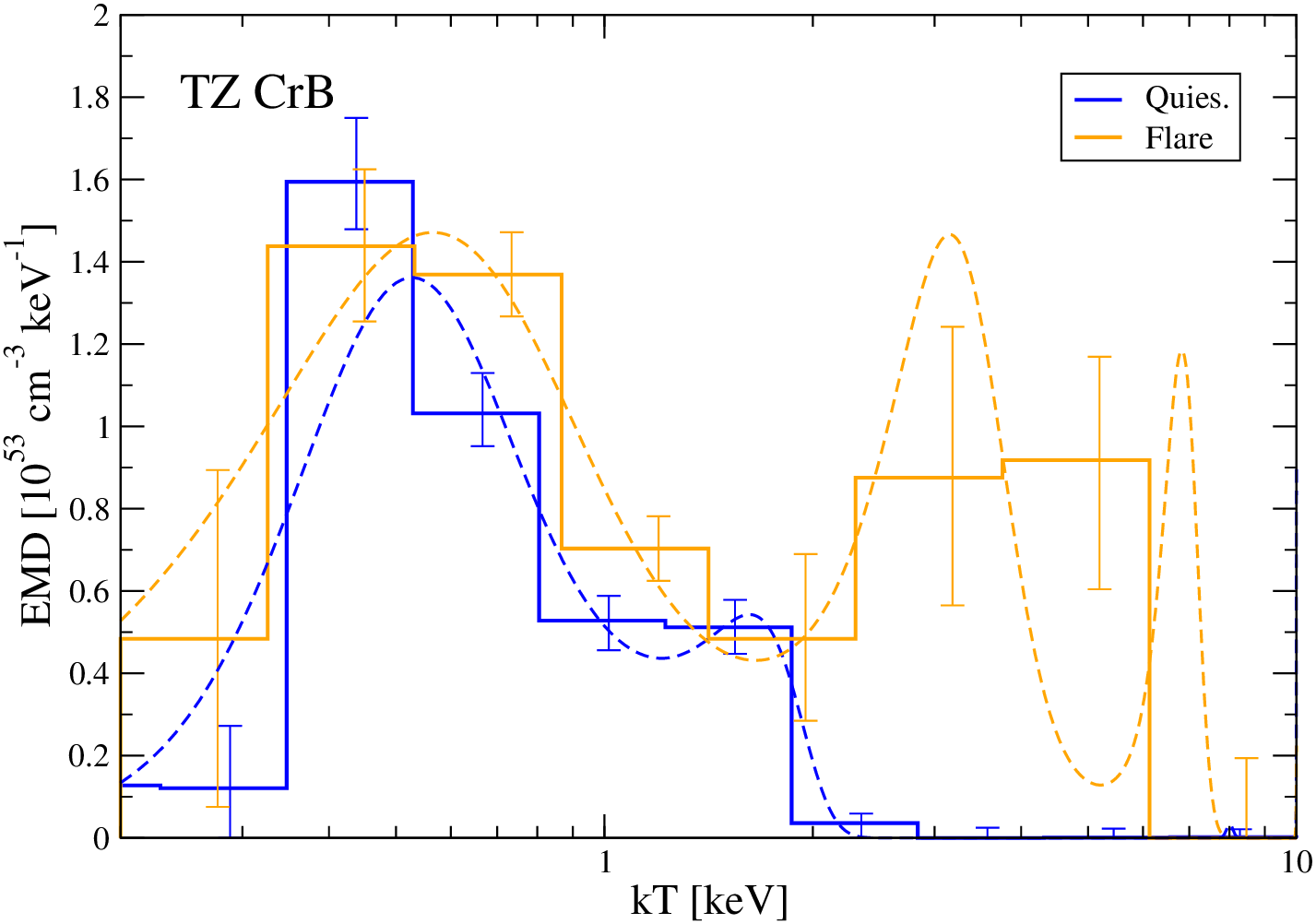}}
    \hspace{25pt}
    \subfigure{\includegraphics[scale=0.35]{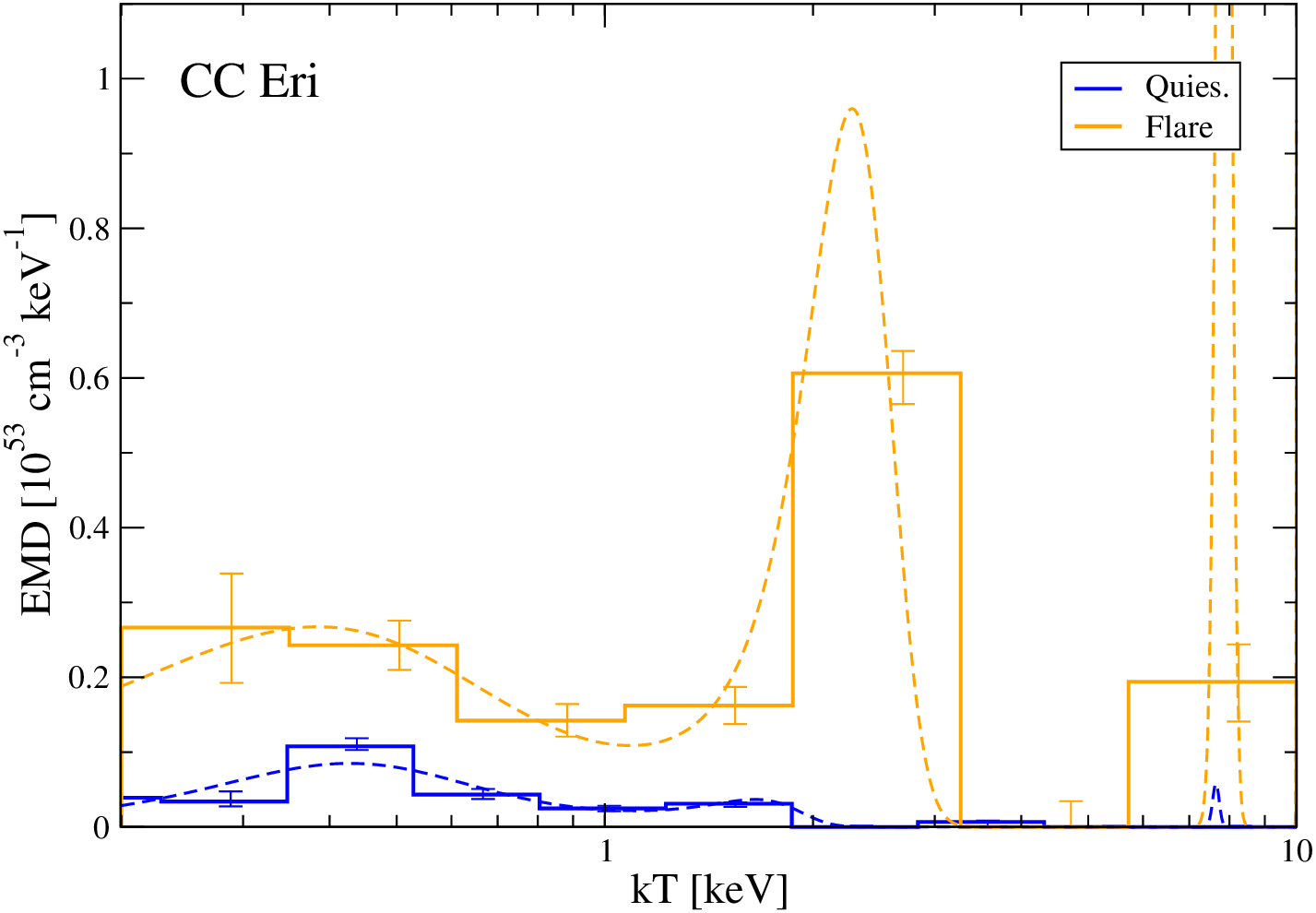}}
    \\
    \caption{Emission measure distributions for flare and quiescence states.
    Solid lines are the binned $EMD$ models with 1$\sigma$ errors. Dotted
    lines are the exponent of polynomials $EMD$ models.
    For scaling purposes only, a solar Fe/H abundance of 3.24$\times10^{-5}$ is assumed
    \citep{Feldman1992}.}
    \label{fg:EMD}
  \end{center}
\end{figure*}

\begin{figure*}[!t]
  \begin{center}
    \subfigure{\includegraphics[scale=0.35]{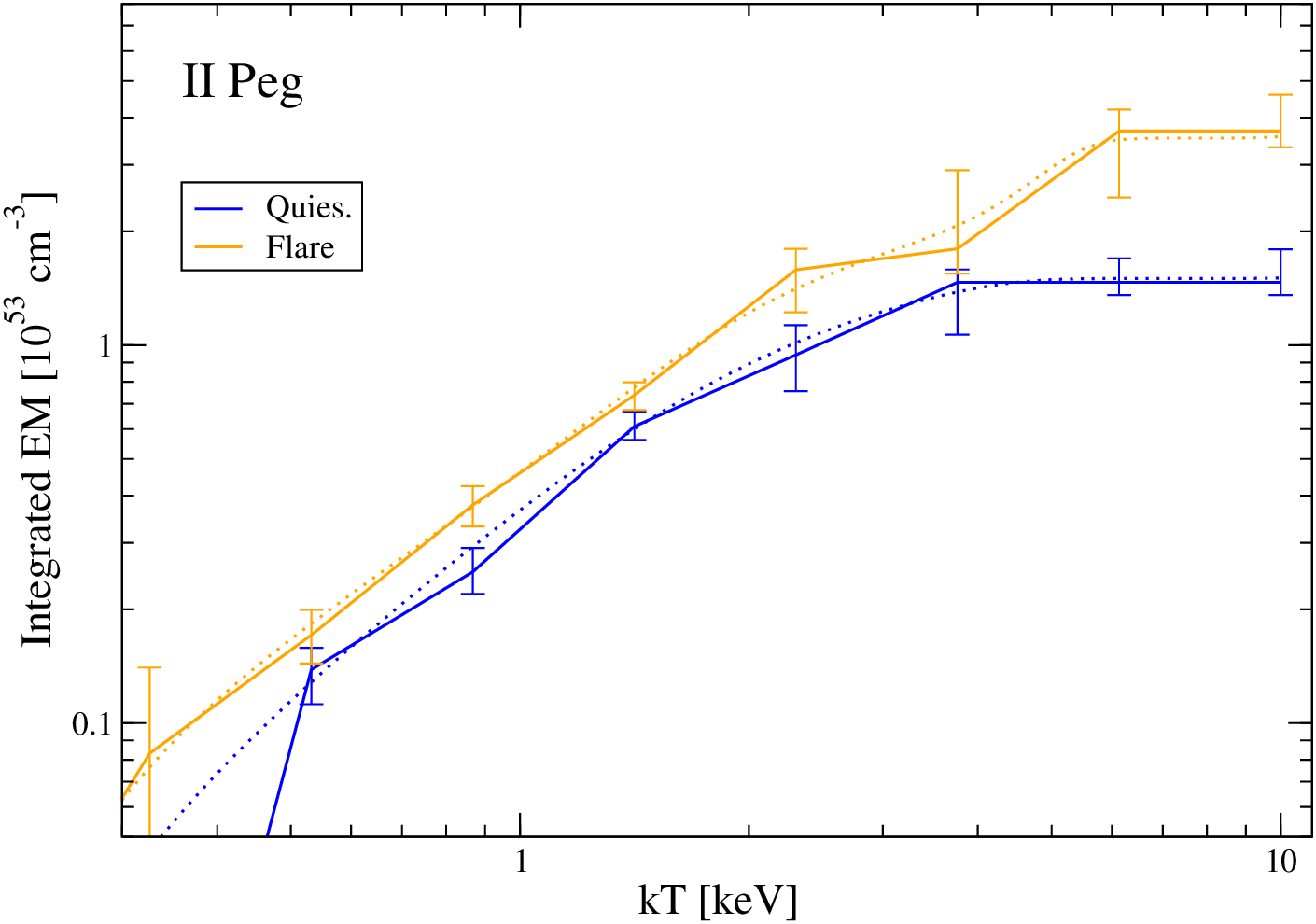}}
    \hspace{25pt}
    \subfigure{\includegraphics[scale=0.35]{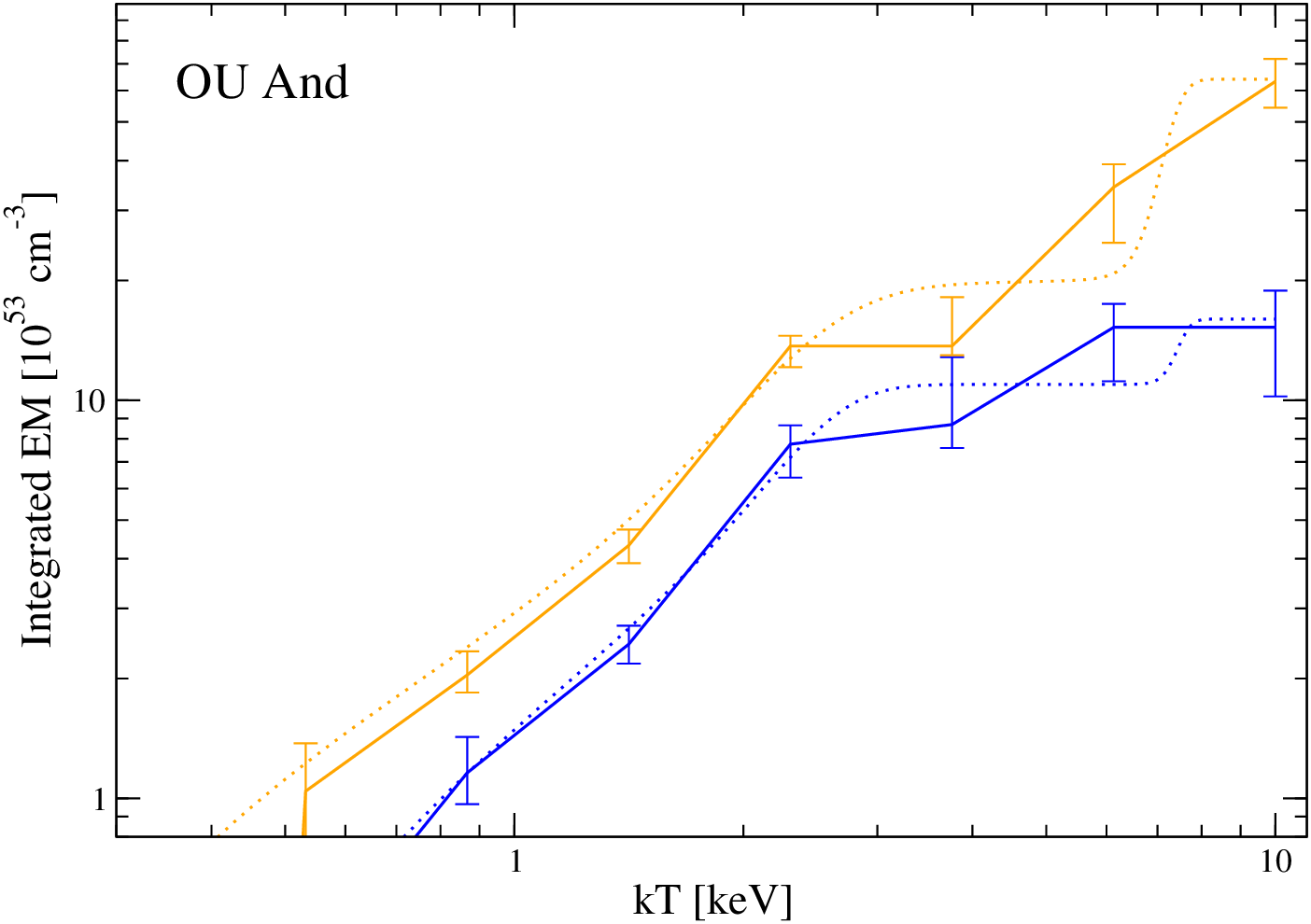}}
    \\
    \vspace{10pt}
    \subfigure{\includegraphics[scale=0.35]{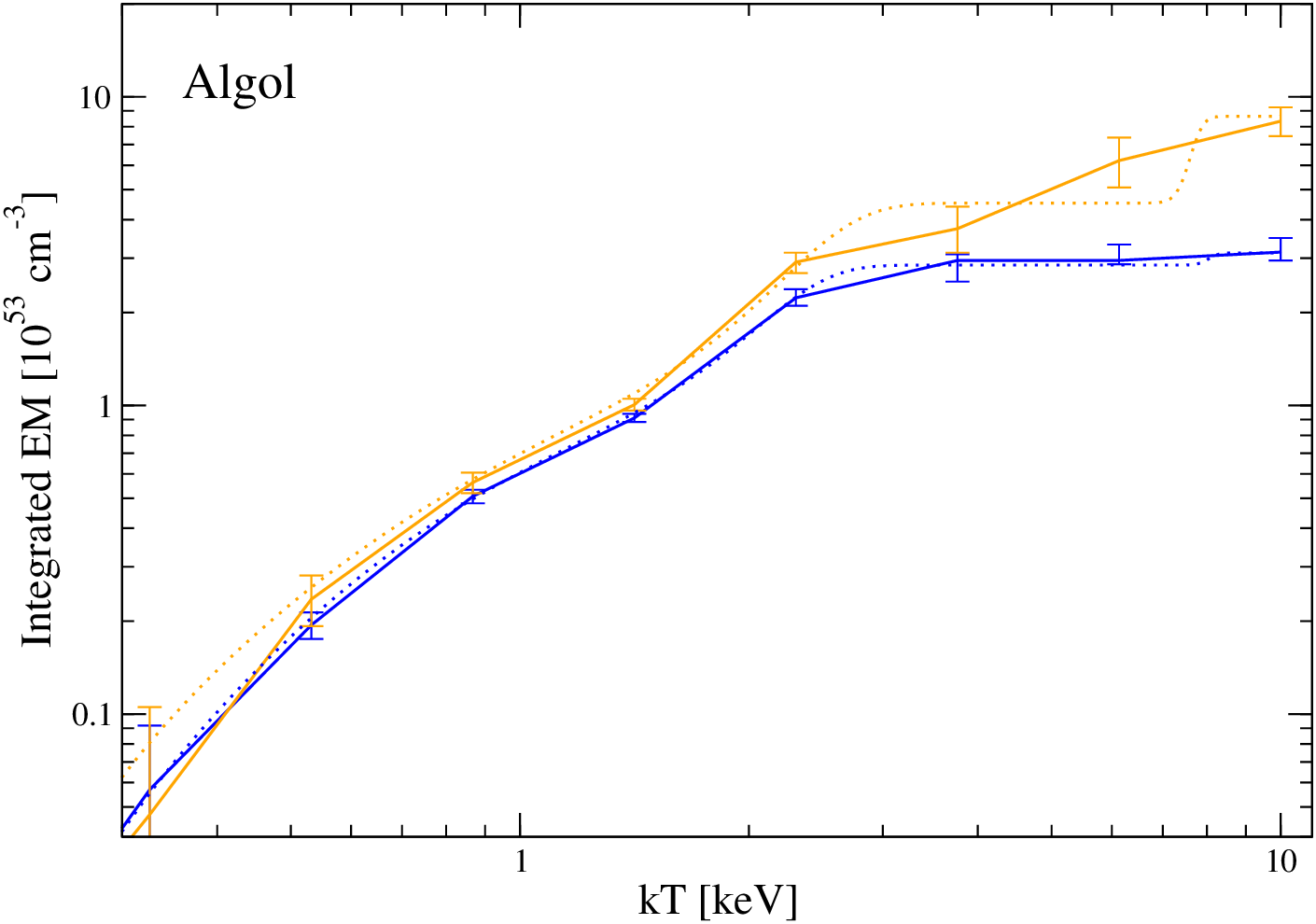}}
    \hspace{25pt}
    \subfigure{\includegraphics[scale=0.35]{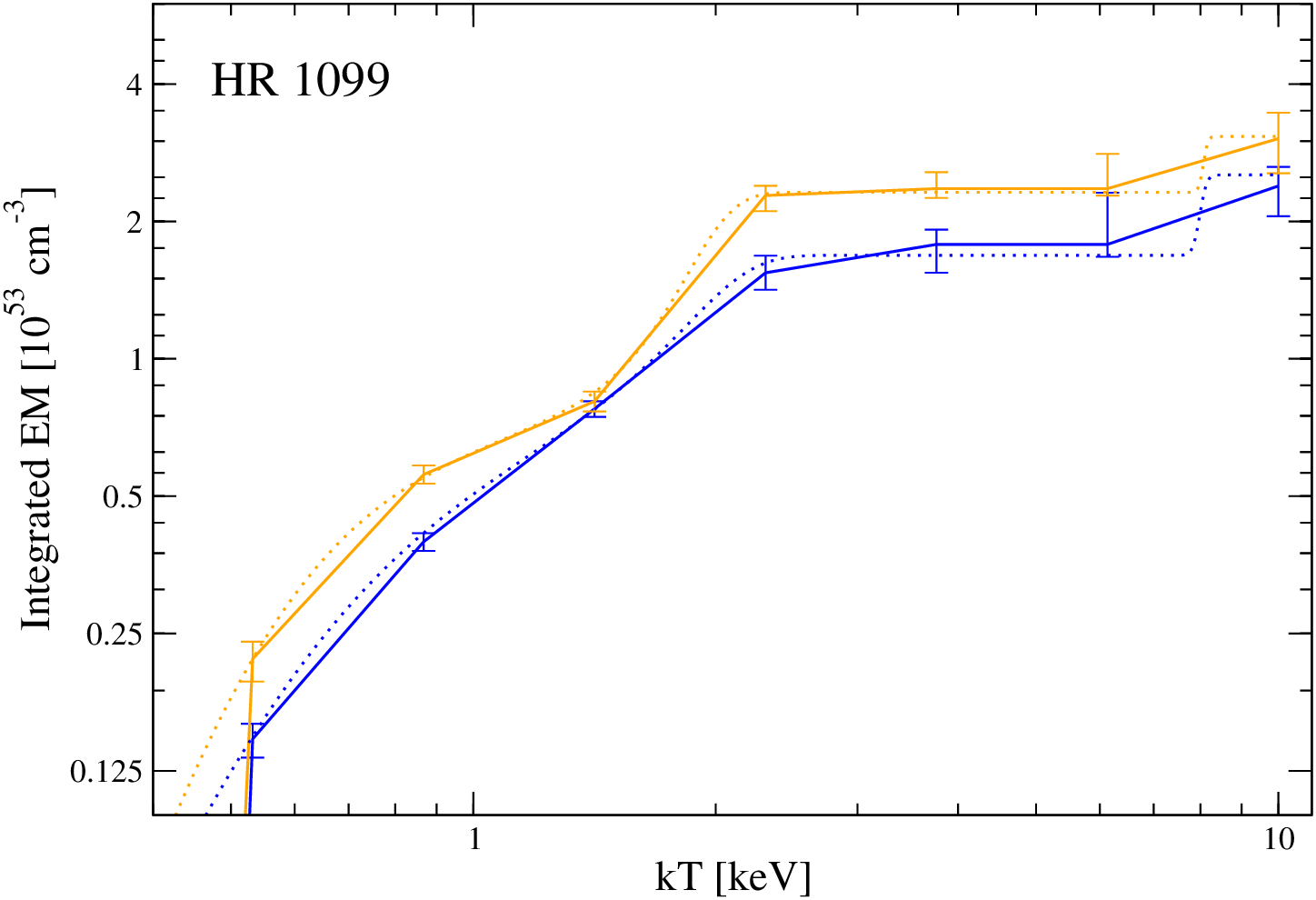}}
    \\
    \vspace{10pt}
    \subfigure{\includegraphics[scale=0.35]{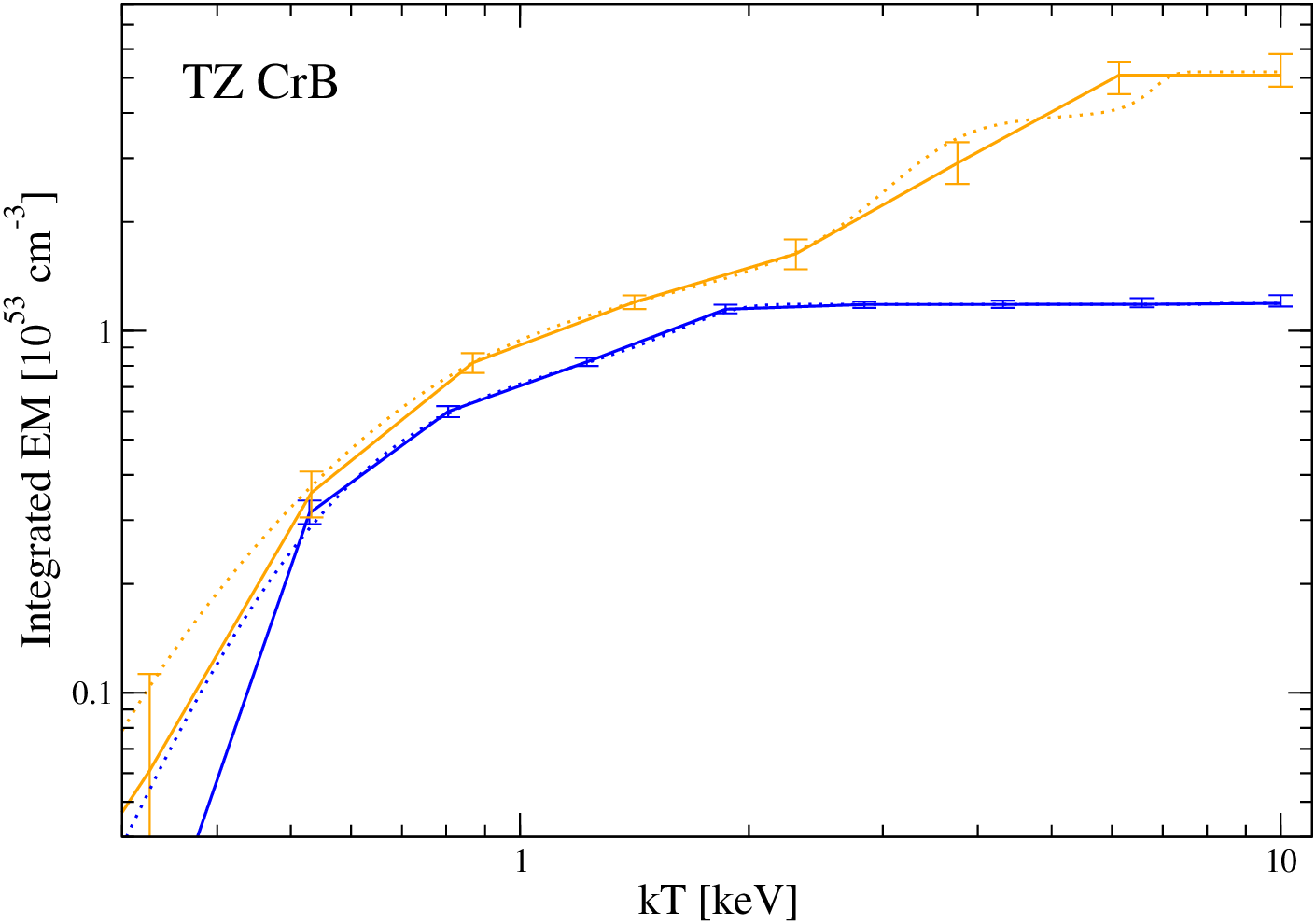}}
    \hspace{25pt}
    \subfigure{\includegraphics[scale=0.35]{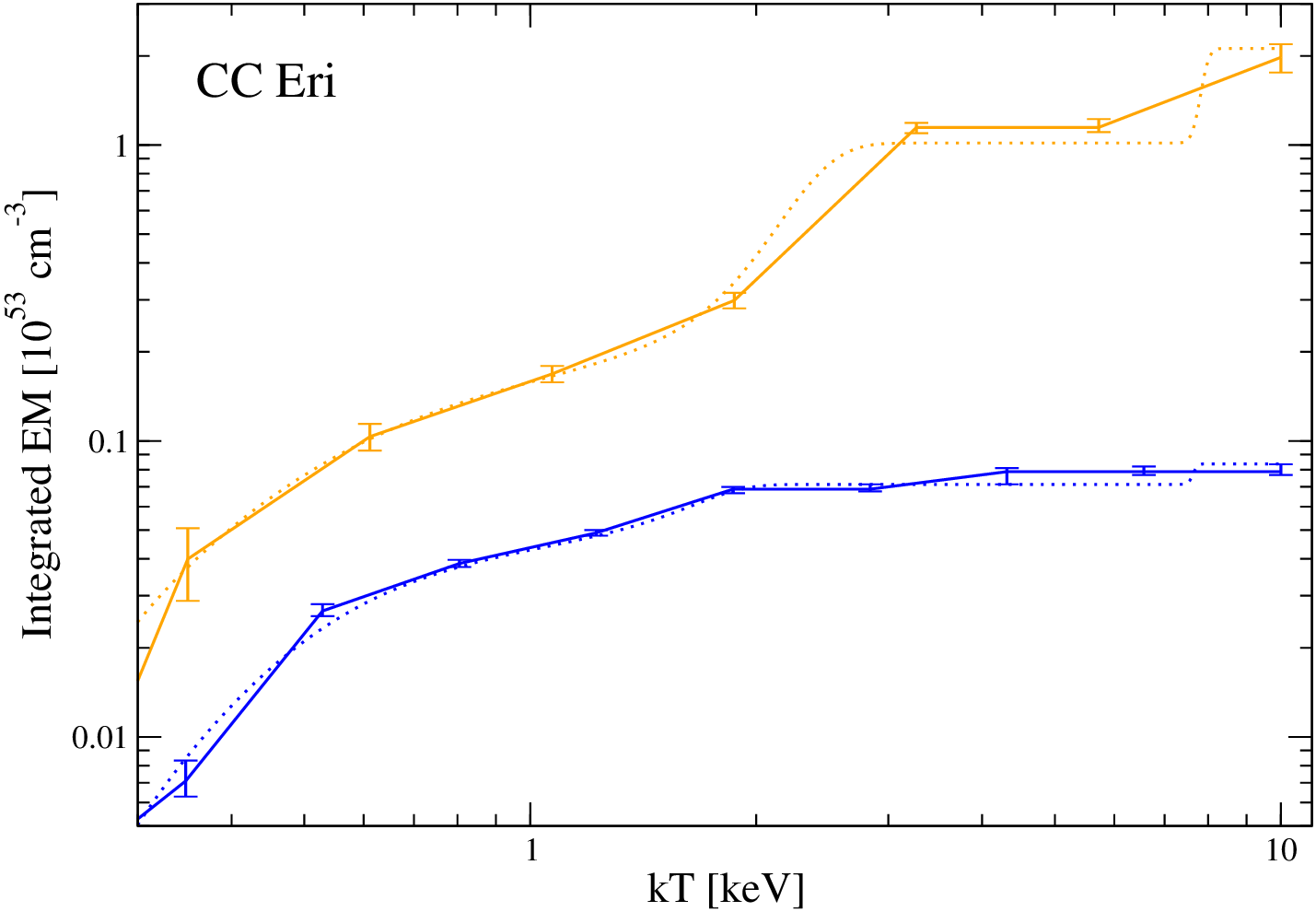}}
    \\
    \caption{The total EM integrated over the $EMD$ from 0.2~keV to $kT$.
    Solid line is from the staircase $EMD$ model and includes 1$\sigma$ errors.
    Dotted line is using the exponent of polynomials $EMD$ model.}
    \label{fg:intg_EM}
  \end{center}
\end{figure*}

The best-fit $EMD$ for each of the six targets is plotted in
figure~\ref{fg:EMD}. The error bars on the staircase $EMD$ model
(eq. (5)) are 1~$\sigma$ errors and include correlation
uncertainties. The smooth dashed line with no confidence intervals
is the exponential model (eq. (6)). Since the $EMD$ is scaled by
the Fe abundance and since we do not measure the absolute Fe/H
abundance, we assume in Fig.~\ref{fg:EMD} a solar abundance of
3.24$\times$10$^{-5}$ \citep{Feldman1992}. This has no effect on
the following results and is done purely for the purpose of
setting a reasonable absolute $EMD$ scale.

In all cases, the two models (staircase and exponential) agree
with each other quite well within the errors over most of the
temperature range considered. The exponential model tends to place
a very sharp peak at $\sim$8~keV, especially for the flaring
states. The reason for this is that the constraints on the $EMD$
in this region are almost exclusively set by Fe XXIV and Fe XXV.
In order to set the observed line flux ratio only one temperature
is needed and the solution tends to a delta function around that
temperature. When we look at the staircase $EMD$ we can see from
the large correlation-induced error bars in the last bins, that
this component can easily be placed in the neighbouring bin
without changing the resulting spectrum significantly.

Comparing the flare and quiescence $EMD$ for each target we see a
general pattern: At temperatures typical of the quiescence state
the $EMD$ is similar or increased by some small factor and at
higher temperatures a new component appears during the flare. In
CC Eri, the lower-$T$ $EMD$ is increased by a considerable factor
of roughly 3. The HR~1099 flare differs from the others by having
very little added $EMD$ at temperatures higher than those of the
quiescence state.

The integrated EM from zero to $kT$ is plotted in
figure~\ref{fg:intg_EM} for all targets with 1$\sigma$ error bars.
The smooth dashed line without confidence intervals is the
integral over the exponential model. In the  0.5-2~keV temperature
region, where many L-shell emissivity curves peak, the IEM from
both models (staircase and exponential) agree very well. At higher
temperatures, the localization of the EM is not as good and the
two models diverge slightly, but they always do re-converge at
higher temperatures, once the integration covers the full range.
At 10~keV the total EM is remarkably similar for both models and
well within the error bars.

In most of the flares (II Peg, OU And, Algol and TZ Crb) the IEM
of the flare, runs parallel to the IEM of quiescence up to
$\sim$2~keV in the log-log plot, which indicates that the thermal
structure remains the same or uniformly increased by a small
factor. In the CC Eri flare, the increase in low-$T$ EM is much
more significant and reaches a factor of 4.3 at 2~keV, where the
flare IEM is still increasing while the quiescence IEM has nearly
reached its maximum value. The HR~1099 flare IEM is generally
parallel and above the quiescence IEM, except around 1.5~keV where
they are similar.

Several theoretical works attempt to predict the $EMD$ of a static
loop or flaring loops. They commonly predict a power law type 
distribution up to a peak temperature:
\begin{equation}
EMD(T) \propto T^{\alpha-1}
\end{equation}
where $\alpha$ is determined by the details of the cooling processes.
Note the different definitions of the EMD in various papers.
For purely radiative cooling, $\alpha = -\gamma+1$ where $\gamma$
is a parameter (in the range $\gamma \approx 0 \pm$0.5) of the 
approximated radiative cooling function:
\begin{equation}
\Lambda(T) \propto T^{\gamma}
\end{equation}
i.e. $\alpha = 0.5-1.5$, while for conductive cooling $\alpha \approx 1.5$ 
and for evaporation $\alpha \approx 0.5$ \citep{Antiochos1980}.
Of special relevance to this work, where we integrate the spectra over
the entire flare, is the calculation by \citet{Mewe1997}.
They predict for a time-averaged $EMD$ of a quasi-statically cooling flaring 
loop $\alpha = 19/8$.
Since we can resolve the $EMD$ only down to 0.2~keV, fitting a 
power law up to the first peak is possible only to in those systems
that peak is around 2~keV. We also fitted the flare 
in the range of 4-10~keV, where the quiescence emission is low. 
The fits were preformed on the IEM curves as the errors there are smaller.
The results are shown in table~\ref{tab:powerlaw}.
We see that $\alpha$ is in the range of 1.5-2, which is slightly higher than expected
from radiative and conductive $EMD$ models, significantly higher than expected from evaporation models and significantly lower than expected from the effect of time averaging 
over a single temperature cooling loop.

\begin{table*}
\caption{\label{tab:powerlaw} Fitted powerlaws for the EMD.}

\begin{tabular}{c | c c c | c c c | c c c}
\hline \hline
       & \multicolumn{3}{|c|}{quiescence (0.2-2 keV)} & \multicolumn{3}{|c|}{Flare (0.2-2 keV)} & \multicolumn{3}{|c}{Flare (4-10 keV)}\\
System & $\alpha$ & $\sigma$ & red. $\chi^{2}$ & $\alpha$ & $\sigma$ & red. $\chi^{2}$ & $\alpha$ & $\sigma$ & red. $\chi^{2}$ \\
\hline
II Peg & 1.47 & 0.14 & 1.33 & 1.48 & 0.18 & 0.06 & 0.8$^{a}$ & 0.4 & 0.26 \\
OU And & 1.96 & 0.24 & 0.51 & 1.87 & 0.16 & 1.4 & 1.5 & 0.3 & 0.2  \\
Algol  & 1.57 & 0.07 & 4.0  & 1.66 & 0.11 & 3.9 & 0.8 & 0.2 & 0.25  \\

\hline
\end{tabular}
\\
$^{a}$ - Due to low EM at very high T, a temperature range of 2.5-6~keV was used.
\end{table*}

\subsection{Abundances}

The abundances relative to Fe, as calculated from the staircase
model are summarized in
Tables~\ref{tab:II_Peg_abund}-\ref{tab:CC_Eri_abund} in the appendix. 
Quoted errors include both flux and $EMD$ uncertainties as explained in
section~\ref{sec:calculating_abundances}. Abundances measured
using the exponential $EMD$ agree well with these results within a
few percent of a std-dev. The difference between the $EMD$
representations is so small that we only quote the values from the
staircase model. Since we obtain the element abundance for
individual ions, elements with more than one ion in the spectrum
may have several abundance values. The abundance value in the
table is the statistically weighted average of these
individual-ion abundances. Note that abundances derived from
different ions of the same element are expected to be consistent,
since this was assumed in the $EMD$ reconstruction algorithm (see
eq.~(\ref{eq:fluxratio})).

\begin{figure*}[!t]
  \begin{center}
    \subfigure{\includegraphics[scale=0.35]{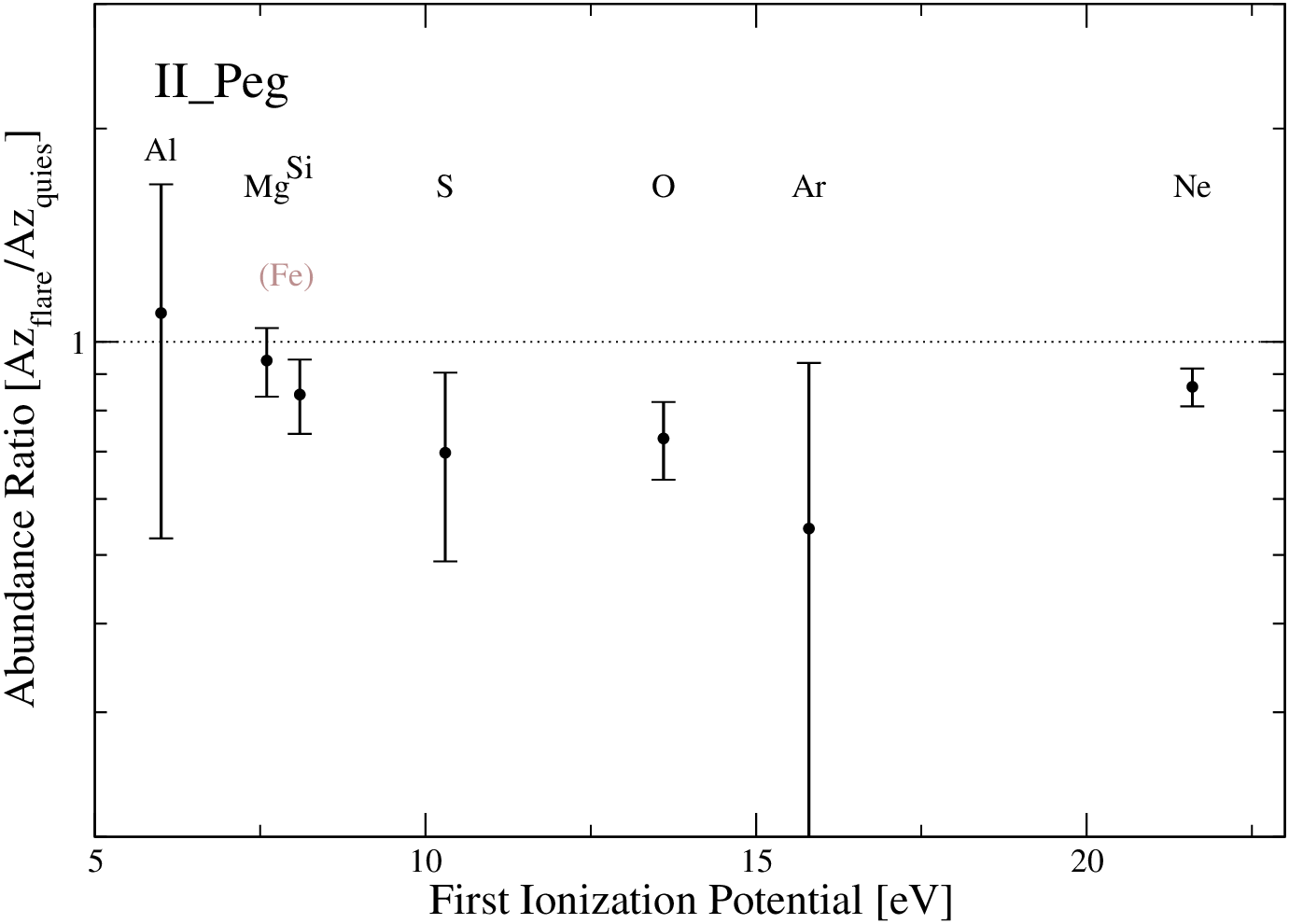}}
    \hspace{25pt}
    \subfigure{\includegraphics[scale=0.35]{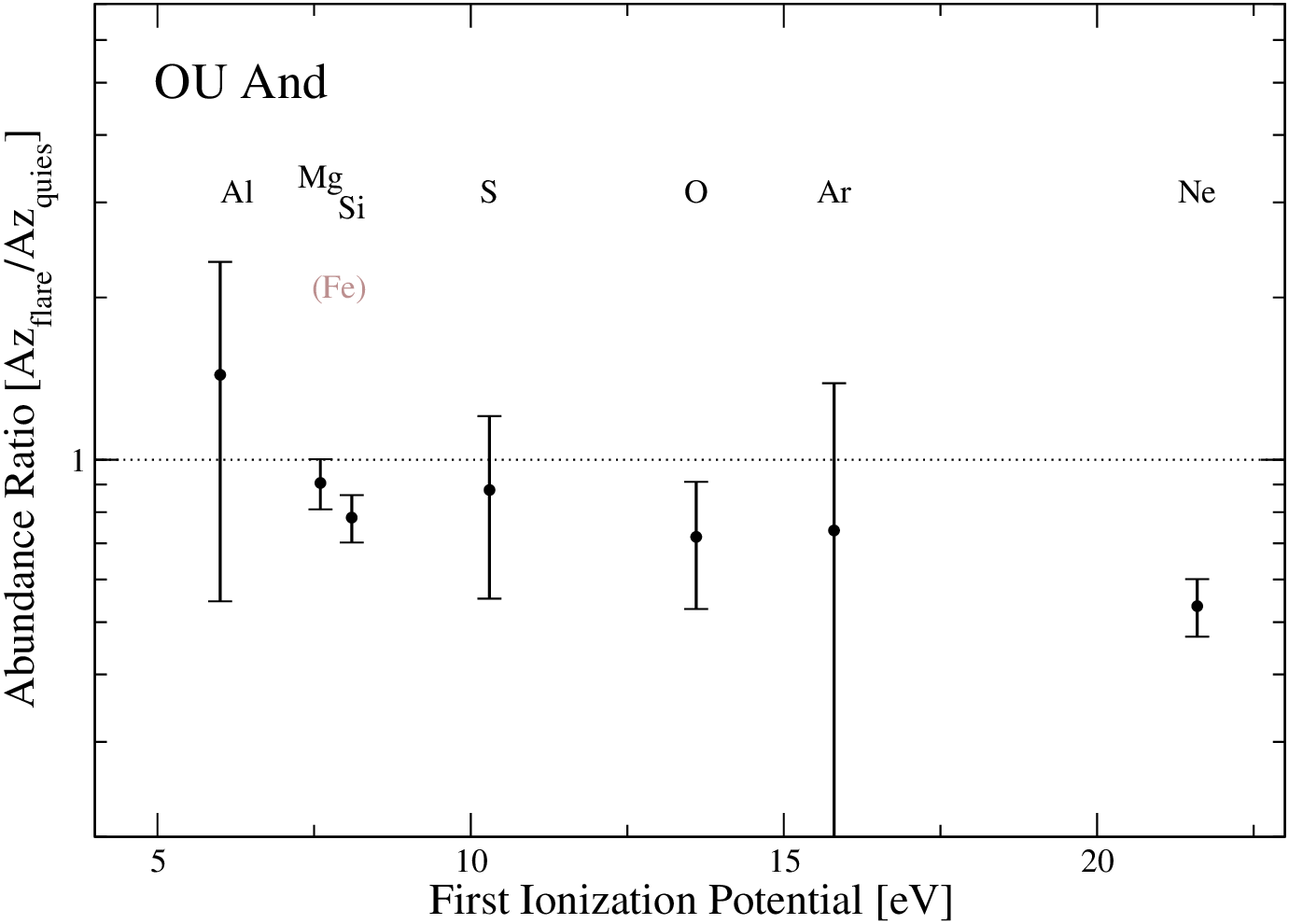}}
    \\
    \vspace{10pt}
    \subfigure{\includegraphics[scale=0.35]{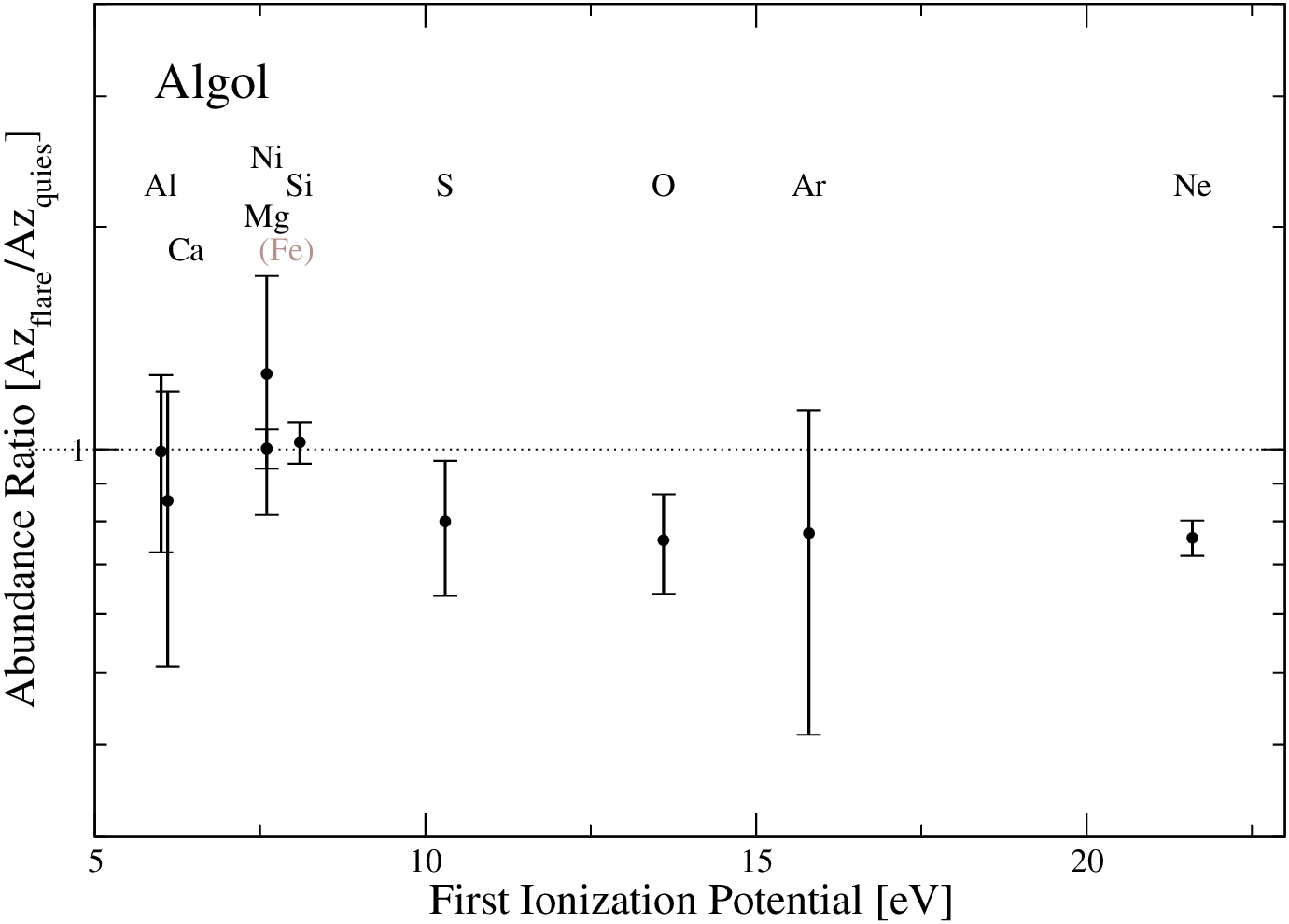}}
    \hspace{25pt}
    \subfigure{\includegraphics[scale=0.35]{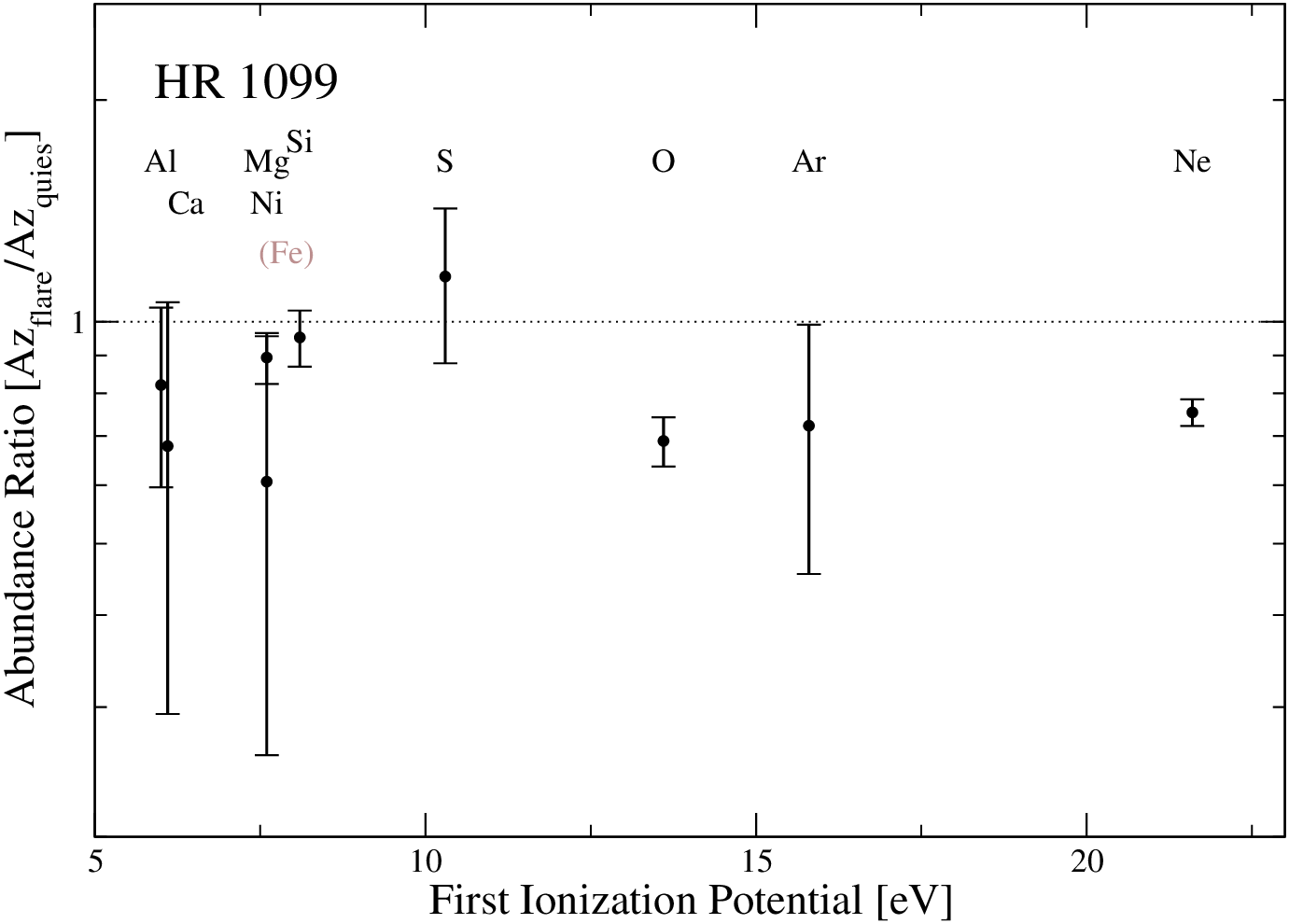}}
    \\
    \vspace{10pt}
    \subfigure{\includegraphics[scale=0.35]{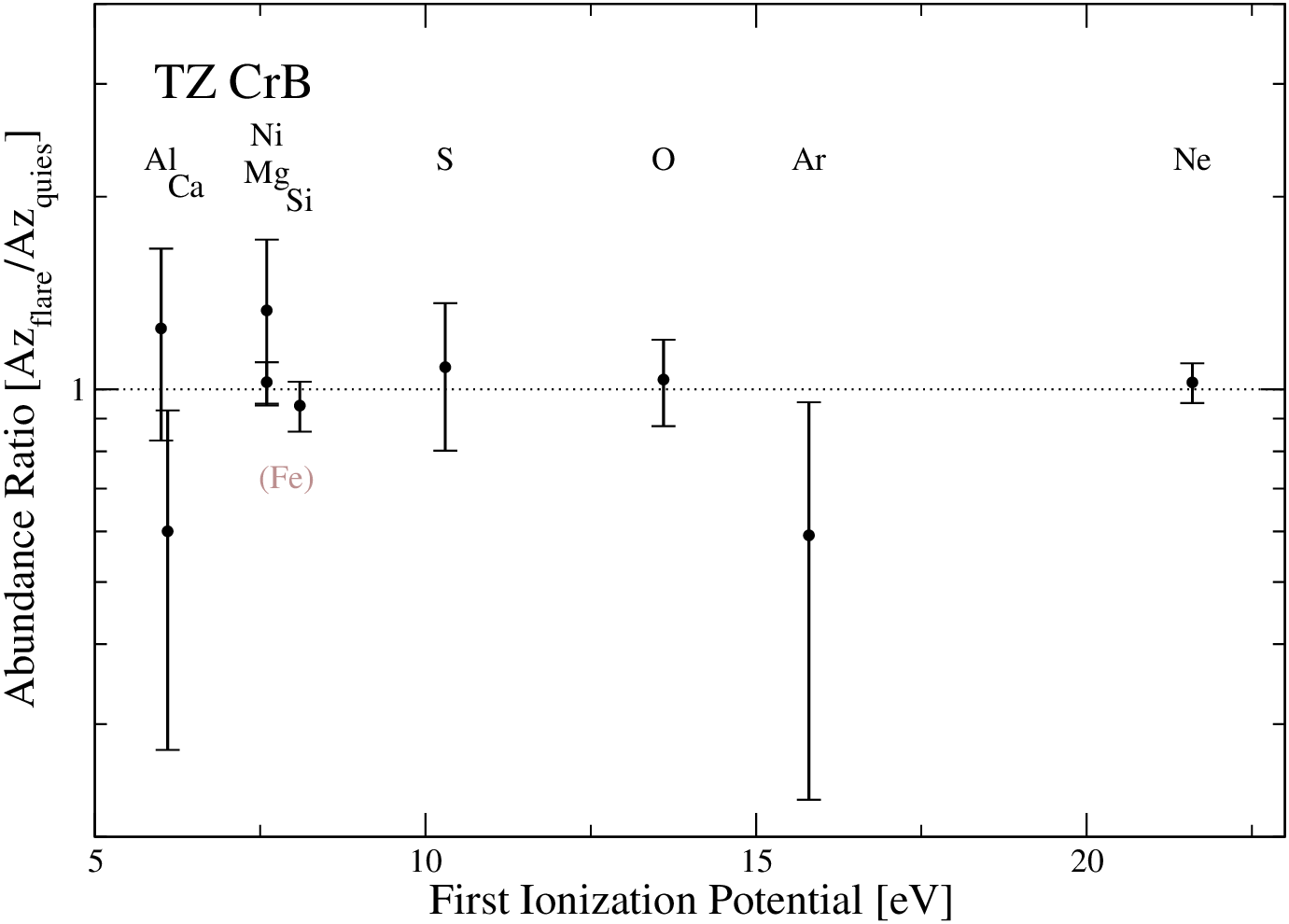}}
    \hspace{25pt}
    \subfigure{\includegraphics[scale=0.35]{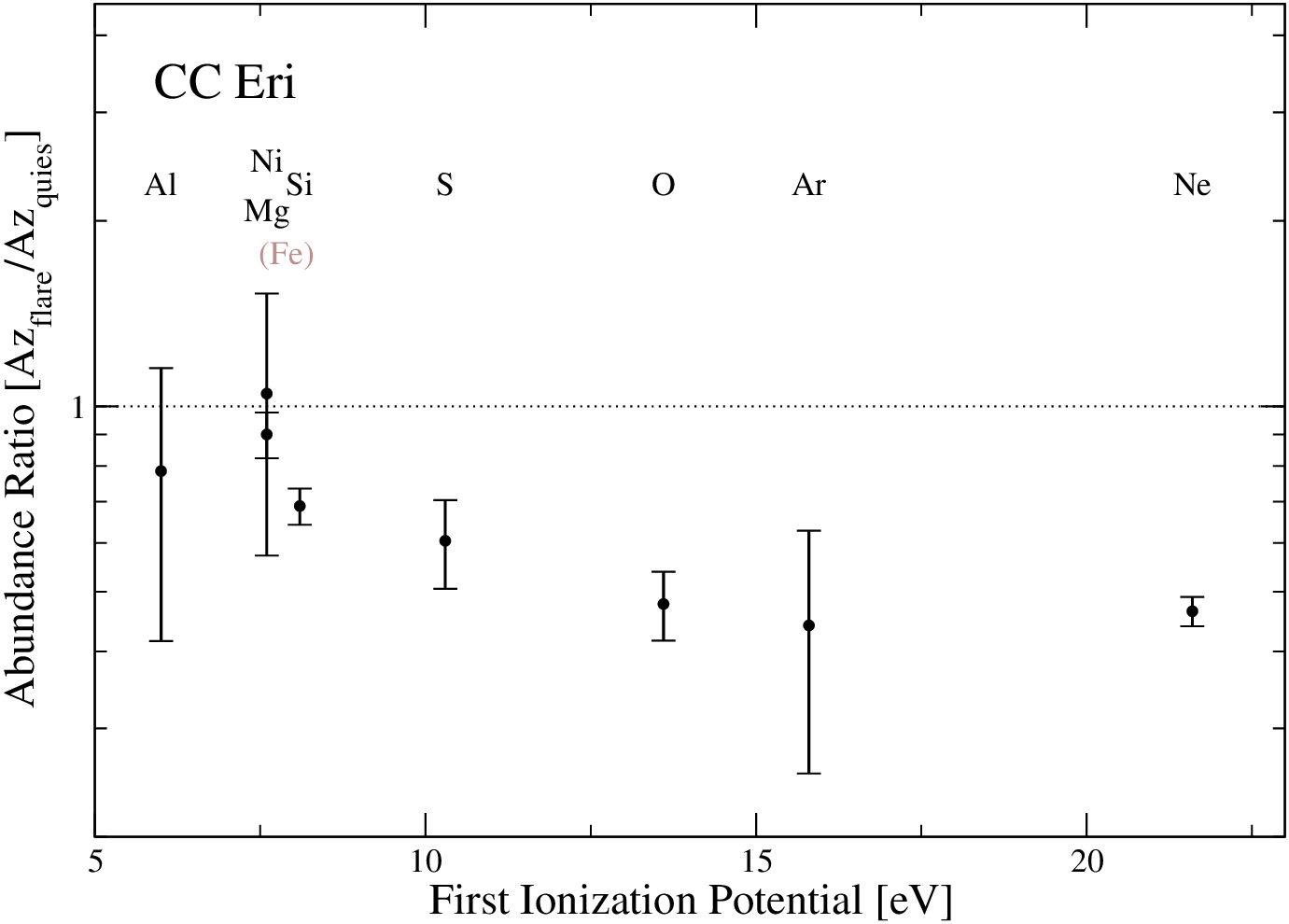}}
    \\
    \caption{Flare abundances relative to quiescence abundances as a function of FIP.}
    \label{fg:AzR}
  \end{center}
\end{figure*}

\section{Discussion}
\subsection{Present work}

The $EMD$ plots depict the average variation in the thermal
structure of the plasma coupled with possible density variations.
The $EMD$ in each temperature bin represents the averaged $EMD$
over the bin temperature range. It is apparent that the systems
that are considered coronally active, namely Algol, HR1099 and OU
And show a similar $EMD$ shape during quiescence: A gradual
increase with $kT$ up to about 2~keV and a sharp drop beyond that,
 becoming negligible at 3-4~keV.
This is in contrast with the two other systems: TZ CrB and CC Eri,
which have most of the emission localized around 0.5~keV during
quiescence, but still have a small amount of EM up to 2~keV. II
Peg is a middle case where the $EMD$ peaks around 1~keV and a
significant portion of the EM is likely to be above 2~keV.

Looking at figure \ref{fg:intg_EM}, for most targets, the flare
and quiescence IEM curves run nearly parallel (in log scale) to
each other up to a point where the EM in the flare increases
sharply relative to quiescence. It is clear that most of the
excess EM is located at the very high temperatures, typically
above 2~keV. HR~1099 is the exception to this rule as no
significant amount of EM is added at high temperatures during the
flare. The IEM in CC Eri is unique. There is much excess low-$T$
EM ($\sim$ factor 4) in the flare, starting apparently from the
lowest temperatures observed.

The similarity in flare and quiescence $EMD$ at temperatures below
2~keV, differing only by a constant factor was also observed by
\citet{Nordon2006} in a flare on $\sigma$ Geminorum. This effect
was clear from a direct comparison of the flare and quiescence
spectra. In that case, the line emission at wavelengths above
12~\AA\ was increased by a uniform 25\%, corresponding to a
uniform $EMD$ increase by the same factor under $\sim$2~keV
temperature. \citet{Audard2001} also reported a similar effect in
a flare on HR~1099 detected by \xmm.

The added EM at low temperatures can be either a real variation of
the quiescence corona, or an effect caused by the time averaging
of the flare heating and/or cooling. In other words, a narrow
temperature component from the flare (moving up or down in $kT$)
will mimic a continuous distribution when averaged over time.
Ideally, we would repeat the $EMD$ analysis for shorter segments
of time to reproduce the time evolution. Unfortunately, that also
means higher statistical uncertainties on the line fluxes, leading
to too high uncertainties on the $EMD$ to make the comparison
feasible. We did extract light curves for the brightest lines in
the spectrum and looked for a direct indication of heating or
cooling. No clear conclusion could be made. In the case of a
narrow temperature component, the cooling through the low-$T$
range would have to be fast enough in order to pass the
significant amount of excess EM from high-T to below the minimum
temperature, without adding much to the time averaged $EMD$. The
fact that this added cool EM is present even in the II Peg
observation, where most of the decay phase is not observed,
suggests that it is not a time averaging effect, but rather a
genuine increase in EM over a broad temperature range in all
flares. 
A note should be added that the $EMD$ is scaled by the
unknown Fe abundance relative to H. Therefore, a possible increase
in Fe abundance could explain the uniform increase in the cool range $EMD$.
This possibility has no effect on, and is consistent with, the FIP related 
abundance variations which we discuss below.

The contrast between the small low-T EM excess and the large high-T EM
excess could be due to rapid expansion (e.g., evaporation, relaxed pinch, loss
of magnetic confinement), which results in a rapid loss of high-T
EM. In this case, the flare either reaches low temperatures when its EM is small, 
or the flare originates in a hot environment and therefore cools down only
to the (high) ambient temperature.

\begin{table*}
\begin{minipage}[!t]{\columnwidth}
\renewcommand{\footnoterule}{}
\caption{\label{tab:total_EM} Total integrated EM$^{a}$ from
0.2~to~10~keV and averaged luminosities$^{b}$
  in the 1.8--20~\AA\ range.}
\begin{tabular}{c c c c c c c c}
\hline \hline
       & \multicolumn{2}{c}{Quiescence} &  \multicolumn{2}{c}{Flare} & EM Ratio & Flare Int. & Total Net Flare \\
       & EM & L$_X$ $^{c}$ & EM & L$_X$ $^{c}$ & Flare/Quies. & Time & Energy \\
Traget & [10$^{53}$~cm$^{-3}$] & [10$^{30}$ erg s$^{-1}$] & [10$^{53}$~cm$^{-3}$] & [10$^{30}$ erg s$^{-1}$] & & [ks] & [10$^{34}$ erg] \\
\hline

II Peg$^{d}$  & 1.47$^{+0.32}_{-0.11}$ & 9.8$\pm$0.2   & 3.7$^{+0.9}_{-0.35}$ & 22.5$\pm$0.2  & 2.5  & 22.7 & 28.7$\pm$0.6   \\
OU And  & 15.25$^{+4.1}_{-5.0}$     & 40.6$\pm$0.8     & 63.2$^{+8.9}_{-9.4}$ & 95.5$\pm$1.3  & 4.1  & 39.5 & 216$\pm$6      \\
Algol   & 3.14$^{+0.37}_{-0.43}$    & 7.69$\pm$0.06    & 8.3$^{+1.2}_{-1.1}$  & 12.8$\pm$0.2  & 2.6  & 13.7 & 7.0$\pm$0.2  \\
HR 1099 & 2.4$^{+0.5}_{-0.3}$       & 9.61$\pm$0.08    & 3.0$^{+0.5}_{-0.5}$  & 10.7$\pm$0.1  & 1.25 & 15.9 & 1.85$\pm$0.2  \\
TZ CrB  & 1.19$^{+0.07}_{-0.03}$    & 2.92$\pm$0.02    & 5.1$^{+0.7}_{-0.6}$  & 9.5$\pm$0.2   & 4.3  & 6.2  & 4.1$\pm$0.1 \\
CC Eri  & 0.079$^{+0.005}_{-0.007}$ & 0.223$\pm$0.003  & 2.0$^{+0.2}_{-0.2}$  & 2.42$\pm$0.04 & 25.3 & 8.0  & 1.76$\pm$0.03 \\

\hline
\end{tabular}

$^{a}$ Using the staircase EMD \\
$^{b}$ Luminosities are averaged over the time intervals as indicated on the light curves\\
$^{c}$ Luminosity of the averaged spectra during quies./flare states\\
$^{d}$ Observation ended while excess flux was still high.

\end{minipage}
\end{table*}

Table~\ref{tab:total_EM} summarizes the values of the total
integrated EM (last data point in figure~\ref{fg:intg_EM}). We see
that in most cases the flares cause an increase in EM by a factor
of 2--4. In HR~1099, interestingly, it increased by only 25\%,
while in CC Eri it increased by a factor of $\sim$25, although the
total EM of CC Eri is still very small relative to the other
systems. This means that the quiescent corona of CC Eri is very
small and the emission in the flare could be attributed entirely
to the flare heated plasma. In the other systems we observe a mix
of flare plasma with a background of the strong quiescent corona.

Overall, there are no dramatic variations in abundances, but
several elements do show a tendency of decreased abundances
relative to Fe during the flares, mainly O and Ne. A similar
effect has also been observed in a flare on $\sigma$~Gem
\citep{Nordon2006}. It should be noted that in those flares where
the low-$T$ EM ($kT <$ 2~keV) during the flare is not
significantly larger than in quiescence, the measured flare
abundances are practically a weighted average of the abundances in
the flaring plasma and the background quiescence plasma.

Figure~\ref{fg:AzR} shows the flare to quiescence abundance ratios
plotted as a function of FIP. Since the abundances are plotted
relative to Fe it means that if the Fe/H abundance has changed in
the flare, all the ratios in the figure will be multiplied by a
uniform factor of $Fe_{quies}/Fe_{flare}$. This may re-scale the
plot, but will not change the pattern or our conclusions. We see
that in all flares, with the exception of TZ CrB, high FIP
elements show reduced abundances. For the low FIP elements the
picture is not as clear as the elements with FIP lower than Fe: Al
and Ca, have large error bars. Still, elements with FIP of 8~eV
and below seem to vary together with Fe (flare/quies is $\sim$1).
This behaviour is similar to the solar FIP effect. TZ CrB is the
exception, with no apparent abundance effects as all abundance
ratios are consistent with unity.

This is somewhat surprising as other works have shown that
increased coronal activity leads to the IFIP effect.
\citet{Audard2003} have examined abundances on RS~CVn systems and
found that high FIP elements were increasingly over-abundant as
the typical temperature (used as an indication of activity)
increased. \citet{Telleschi2005} have found a similar effect in
solar like stars where the FIP effect switched to IFIP as activity
increased. For stars with an IFIP corona, chromospheric
composition would appear to be FIP in comparison. The change in
abundances during the flare indicates that the excess plasma is
not heated coronal plasma, but more likely heated chromospheric
plasma. This would mean that, if flares are at all responsible for
the IFIP effect, the selective element transport would have to
operate during the cooling or post-flare stage since we detect a
low-FIP enriched composition during the flare itself.

From looking at the light curves in figure~\ref{fg:lc},
we clearly see two types of flare behaviours: Symmetric flares
with a sharp peak and a rapid decay (CC Eri \& Tz CrB), and
asymmetric flares that rise fast, but decay slowly, which creates
a broad peak (II Peg, HR1099, OU And). Algol, at first glance,
appears to belong to the first kind, but the flare occurred just
as the system was coming out of eclipse. As noted by
\citet{Chung2004}, it is likely that if most of the emission
originates from near chromospheric level, the rise phase in the X-ray light
curve is governed by the exposure of the flaring region behind the
eclipsing companion and not by the rise of the flare. 
Both of the rapid decay flares occurred on the cooler dwarf stars, 
while the long duration flares are all on the giant or sub-giant stars.
However, CC Eri flare shows the clearest FIP effect in the flare relative 
to quiescence, similar to the long duration flares, while TZ CrB shows 
no abundance variations. The sample is too small to draw any substantiated
conclusions, but it seems that there is no clear correlation between the type 
of flare and abundance effects.

\subsection{Comparison With Previous Works}
\label{sec:Prev_works} Several of the observations used in this
work (table~\ref{tab:obs_table}) were published before by
other researchers, although for the most part the focus was not on
the flare effects. Also, the methods were different than ours and
the presentation of results makes comparisons difficult. We bring
here a short summary of the results of those individual works as
well as references to other relevant high-resolution X-ray
observations of these targets.

Obs. 1451 of II Peg was analyzed by \citet{Huenemoerder2001} who
found that during the flare, at temperatures of $\log T = 7.3$ to
$\log T = 8.0$~K, a large EM component was added while the cooler
emission was hardly changed. We find a similar pattern, but the
cooler emission tends to be slightly higher during the flare, 
which could be due to Fe abundance variation as previously discussed. 
The abundance ratio of Ne/Fe relative to solar drops from 22$\pm$6
preflare to 17$\pm$5 (a factor of 0.77) during the flare, which is
only slightly more than the present ratio of 0.86$\pm$0.05.

Obs. 1892 of OU And has not been fully analyzed to date. The
system has also been observed by {\it XMM-Newton} during a
quiescent state as reported by \citet{Gondoin2003}.

Obs. 604 of Algol was analyzed by \citet{Chung2004}. They
investigated slight line shifts and showed that the X-ray emission
is dominated by the secondary star in the system. The corona is
likely to be asymmetric and located closer toward the center of
mass of the system. There is no special treatment of the flare.
Algol is a well studied target in quiescence by the high
resolution instruments. \citet{Schmitt2004} used a line fitting
method on a different \cxc\ LETG observation, to get a smooth
$EMD$ solution for Algol in its quiescence state. Depending on the
details of the fitting, they get various $EMD$ solutions with
peaks at 1--2~keV, consistent with ours. A flare on Algol has also
been observed during an eclipse by {\it XMM-Newton} and allowed
for spatial information to be extracted, see: \citet{Schmitt2003}.
Only limited spectral information was used.

Obs. 62538 of HR 1099 was performed as part of a multi-wavelength
campaign involving {\it Chandra}, {\it EUVE}, {\it HST} and the
{\it VLA}. Results from this collaboration were published by
\citet{Ayres2001}, however no $EMD$ or abundances were published.
HR 1099 was also observed during a large flare with {\it
XMM-Newton} as reported by \citet{Audard2001} who found an
abundance enhancement of the low-FIP elements Fe and Si during the
flare, while high-FIP Ne remained constant. (The quiescent
abundances were reported to have an overall IFIP trend.) This FIP
bias during the flare is similar to what we find for HR~1099 and
in general for five of the six flares we have analyzed.
\citet{Osten2004} used EUVE observations and found an HR~1099
quiescence $EMD$ in agreement with the one presented here, showing
a broadly distributed EM in the range of 0.4~keV to 2.5~keV where
their plot is cut-off.

Obs. 15 of TZ CrB which was part of a campaign that included {\it
Chandra}, {\it EUVE}, and the {\it VLA}, was analyzed by
\citet{Osten2003}. They report an overall increase in absolute
abundances and no FIP related pattern during either quiescence or
flares. In our work absolute abundances (relative to H) were not
calculated, but we also find no FIP pattern in the flare relative
to quiescence. We also detect a Ni XIX line at 12.43~$\AA$, while
\citet{Osten2003} claim the absence of Ni XIX lines. Though this
Ni line is somewhat blended with Fe lines, the combined emissivity
of Fe contributes less than 50\%\ of the total flux in the feature
according to our flux measurement method. An $EMD$ was constructed
by \citet{Osten2003}, but no error bars are given, making a
comparison difficult. Their $EMD$, similar to ours, features two
major components, one at 0.5~keV which exists also in quiescence
and the other at about 3~keV which emerges with the flare.
This target has also been observed with {\it XMM-Newton} by \citet{Suh2005}.
They confirm the absence of a clear FIP bias.

Obs. 6132 and 4513 of CC Eri were not published yet. CC Eri was
previously observed in low resolution by ASCA, simultaneously with
EUVE and the VLA \citep{Osten2002}. No flares were observed.

\section{Conclusions}

We have analyzed six large X-ray flares observed with \cxc\ on six
different systems. Emission measure distribution and integrated EM
were calculated, including well-localized confidence intervals
allowing for an unambiguous comparison between flare and
quiescence states. Relative abundances were measured from the
$EMD$ and compared between states. We verified that the derived
abundances do not depend on the representation of the $EMD$.
Integration over both staircase and exponential $EMD$s produce the
same total EM to high accuracy, showing that the inevitable
degeneracy of the $EMD$ solution has little effect on the total
EM. Thus, we conclude that in contrast with the local $EMD(T)$,
the EM integrated from a low temperature to a (continuously
varying) high temperature is a convenient quantity useful for
comparison between different emission states and different
targets.

During the six flares analyzed in this work, the $EMD$ at
temperatures below $\sim$2keV appears to be similar to that during
quiescence with a small, roughly uniform enhancement observed
during the flare. In five of the six flares, the added EM is
predominantly at temperatures of $kT>2$~keV. 
The total EM is increased by a factor of 2--4
for most flares, but by a factor of $\sim$25 for the CC Eri flare
and only by 25\% for HR~1099. Five out of the six targets show a
statistically significant flare FIP bias in which the high-FIP to
low-FIP abundance ratios decrease during the flares. The exception
is the TZ CrB flare that showed no statistically significant
abundance variations. We conclude that in our sample, flaring
activity tends to evaporate plasma with abundances biased toward a
(solar) FIP effect and not an IFIP effect. Note that this
conclusion does not require the knowledge of and therefore is
independent of photospheric or quiescent coronal abundances.

The different element composition observed during the flare and
the significant increase in total EM indicate that the flaring
plasma is likely to be heated chromospheric plasma, rather than
locally heated coronal plasma. This might support the
chromospheric evaporation scenario. On the other hand, our results
are inconsistent with previous suggestions that flaring activity
and chromospheric evaporation produce an IFIP effect. If flares
are indeed responsible for the transition from FIP to IFIP coronal
composition in very active coronae, and in order to be consistent 
with our observations, the fractionation mechanism must operate during 
the cooling stage or after the flare.
If what we observe in flares is indeed photospheric composition 
plasma heated to coronal temperatures, flares may provide a mean
to measure photospheric abundances in active stars.
On the other hand, it could be the flare itself that introduces the FIP
bias by selectively injecting low FIP material into the corona.

\acknowledgements The research at the Technion was supported by
ISF grant 28/03 and by a grant from the Asher Space Research
Institute.


\Online

\begin{appendix}

\section{Measured Line Fluxes}

\begin{table*}[!h]

\caption{\label{tab:line_fluxes1} Measured line fluxes in
10$^{-4}$ photons s$^{-1}$ cm$^{-2}$ of II Peg and OU And} 

\begin{tabular}{lc | cccccc | cccccc |}

\hline \hline
      &       & \multicolumn{6}{|c|}{{\bf II Peg}}          &\multicolumn{6}{|c|}{{\bf OU And}}   \tabularnewline
      &       & \multicolumn{2}{|c}{Quiescence} & \multicolumn{2}{c}{Flare} & \multicolumn{2}{c|}{Flare/Quies.}
              & \multicolumn{2}{|c}{Quiescence} & \multicolumn{2}{c}{Flare} & \multicolumn{2}{c|}{Flare/Quies.} \tabularnewline
Ion   & Wave  & Flux   & Err & Flux & Err & Ratio & Err   & Flux & Err & Flux & Err & Ratio & Err   \\
\hline

Fe XVII & 15.01 & 1.18 & 0.17 & 1.45 & 0.17 & 1.23 & 0.22 & 0.48 & 0.08 & 1.00 & 0.13 & 2.09 & 0.43 \\
Fe XVIII & 14.21 & 0.64 & 0.15 & 0.83 & 0.16 & 1.30 & 0.39 & 0.27 & 0.06 & 0.43 & 0.10 & 1.63 & 0.52 \\
Fe XIX & 13.51 & 0.54 & 0.14 & 0.72 & 0.15 & 1.35 & 0.45 & 0.26 & 0.06 & 0.40 & 0.09 & 1.55 & 0.50 \\
Fe XX & 12.84 & 0.47 & 0.13 & 0.68 & 0.15 & 1.43 & 0.49 & 0.27 & 0.06 & 0.40 & 0.08 & 1.47 & 0.43 \\
Fe XXI & 12.28 & 0.69 & 0.11 & 0.80 & 0.12 & 1.16 & 0.26 & 0.36 & 0.05 & 0.61 & 0.08 & 1.70 & 0.31 \\
Fe XXII & 11.77 & 0.50 & 0.09 & 0.66 & 0.10 & 1.31 & 0.29 & 0.23 & 0.03 & 0.43 & 0.05 & 1.88 & 0.36 \\
Fe XXIII & 11.00 & 0.42 & 0.07 & 0.59 & 0.07 & 1.40 & 0.28 & 0.32 & 0.03 & 0.55 & 0.05 & 1.71 & 0.21 \\
Fe XXIV & 10.64 & 0.50 & 0.07 & 1.06 & 0.09 & 2.12 & 0.35 & 0.50 & 0.03 & 1.18 & 0.06 & 2.34 & 0.19 \\
Fe XXV & 1.85 & 0.04 & 0.14 & 0.38 & 0.17 & 10.13 & 37.70 & 0.16 & 0.07 & 0.83 & 0.13 & 5.09 & 2.21 \\
Fe XXVI & 1.78 & 0.00 & 0.27 & 0.15 & 0.28 & - & - & 0.03 & 0.13 & 0.30 & 0.21 & 9.58 & 40.17 \\
O VII & 21.60 & 3.26 & 2.20 & 2.36 & 2.03 & 0.72 & 0.79 & 0.05 & 0.77 & 0.00 & 1.11 & 0.01 & 20.65 \\
O VIII & 18.97 & 16.75 & 1.00 & 20.32 & 1.02 & 1.21 & 0.09 & 1.57 & 0.25 & 2.31 & 0.40 & 1.47 & 0.35 \\
Ne IX & 13.45 & 3.40 & 0.23 & 3.96 & 0.23 & 1.17 & 0.10 & 0.17 & 0.05 & 0.16 & 0.07 & 0.94 & 0.50 \\
Ne X & 12.13 & 10.73 & 0.32 & 13.83 & 0.33 & 1.29 & 0.05 & 0.96 & 0.07 & 1.16 & 0.10 & 1.21 & 0.14 \\
Mg XI & 9.17 & 0.42 & 0.05 & 0.48 & 0.06 & 1.16 & 0.21 & 0.13 & 0.02 & 0.22 & 0.03 & 1.73 & 0.37 \\
Mg XII & 8.42 & 0.76 & 0.07 & 1.23 & 0.08 & 1.62 & 0.18 & 0.40 & 0.03 & 0.80 & 0.05 & 2.02 & 0.22 \\
Al XII & 7.76 & 0.04 & 0.04 & 0.00 & 0.04 & 0.00 & 1.13 & 0.007 & 0.014 & 0.01 & 0.02 & 1.45 & 4.27 \\
Al XIII & 7.17 & 0.09 & 0.04 & 0.21 & 0.06 & 2.33 & 1.25 & 0.029 & 0.016 & 0.10 & 0.03 & 3.38 & 2.13 \\
Si XIII & 6.65 & 0.53 & 0.06 & 0.60 & 0.06 & 1.13 & 0.17 & 0.23 & 0.02 & 0.32 & 0.03 & 1.42 & 0.20 \\
Si XIV & 6.18 & 0.57 & 0.07 & 0.99 & 0.08 & 1.73 & 0.25 & 0.35 & 0.03 & 0.69 & 0.05 & 1.97 & 0.21 \\
S XV & 5.04 & 0.31 & 0.10 & 0.35 & 0.12 & 1.14 & 0.54 & 0.068 & 0.036 & 0.19 & 0.06 & 2.85 & 1.77 \\
S XVI & 4.73 & 0.45 & 0.12 & 0.63 & 0.14 & 1.40 & 0.47 & 0.13 & 0.04 & 0.25 & 0.07 & 1.95 & 0.86 \\
Ar XVII & 3.95 & 0.140 & 0.073 & 0.130 & 0.087 & 0.93 & 0.79 & 0.046 & 0.028 & 0.053 & 0.047 & 1.15 & 1.25 \\
Ar XVIII & 3.73 & 0.061 & 0.076 & 0.132 & 0.098 & 2.17 & 3.16 & 0.011 & 0.031 & 0.081 & 0.055 & 7.10 & 19.59 \\
Ca XIX & 3.18 & 0.000 & 0.060 & 0.131 & 0.082 & - & - & 0.017 & 0.027 & 0.061 & 0.046 & 3.67 & 6.53 \\
Ca XX & 3.02 & 0.000 & 0.079 & 0.073 & 0.110 & - & - & 0.019 & 0.031 & 0.117 & 0.056 & 6.20 & 10.63 \\
Ni XIX & 12.43 & 0.031 & 0.073 & 0.091 & 0.086 & 2.98 & 7.66 & 0.000 & 0.028 & 0.045 & 0.046 & - & - \\

\hline \multicolumn{14}{l}{{\it NOTE:} Fluxes of H-like ion lines
include both transitions of the
unresolved Ly-$\alpha$ doublet.}\\
\multicolumn{14}{l}{Fluxes of He-like ion lines include only the resonant transition.}\\
\multicolumn{14}{l}{Fluxes of L-shell Fe ion lines include all
transitions within $\pm$0.03~\AA\ of the specified wavelength.}

\end{tabular}
\end{table*}

\begin{table*}[!h]

\caption{\label{tab:line_fluxes2} Measured line fluxes in
10$^{-4}$ photons s$^{-1}$ cm$^{-2}$ of Algol and HR 1099}

\begin{tabular}{lc | cccccc | cccccc |}

\hline \hline
    &      &\multicolumn{6}{|c|}{{\bf Algol}}                 &\multicolumn{6}{|c|}{{\bf HR 1099}} \tabularnewline
    &      &\multicolumn{2}{|c}{Quiescence} & \multicolumn{2}{c}{Flare} & \multicolumn{2}{c|}{Flare/Quies.}
           &\multicolumn{2}{|c}{Quiescence} & \multicolumn{2}{c}{Flare} & \multicolumn{2}{c|}{Flare/Quies.} \tabularnewline
Ion & Wave & Flux   & Err & Flux & Err & Ratio & Err   & Flux & Err & Flux & Err & Ratio & Err   \\

\hline

Fe XVII & 15.01 & 4.77 & 0.18 & 5.55 & 0.35 & 1.17 & 0.09 & 4.12 & 0.22 & 5.73 & 0.31 & 1.39 & 0.11 \\
Fe XVIII & 14.21 & 2.67 & 0.15 & 2.73 & 0.28 & 1.02 & 0.12 & 2.09 & 0.19 & 2.49 & 0.29 & 1.19 & 0.17 \\
Fe XIX & 13.51 & 2.33 & 0.15 & 2.50 & 0.28 & 1.07 & 0.14 & 1.70 & 0.16 & 1.93 & 0.22 & 1.14 & 0.17 \\
Fe XX & 12.84 & 2.12 & 0.14 & 2.39 & 0.27 & 1.13 & 0.15 & 1.78 & 0.22 & 1.71 & 0.21 & 0.96 & 0.17 \\
Fe XXI & 12.28 & 2.25 & 0.13 & 2.73 & 0.24 & 1.21 & 0.13 & 1.94 & 0.14 & 1.95 & 0.19 & 1.00 & 0.12 \\
Fe XXII & 11.77 & 1.82 & 0.09 & 2.04 & 0.17 & 1.12 & 0.11 & 1.18 & 0.10 & 1.35 & 0.14 & 1.14 & 0.15 \\
Fe XXIII & 11.00 & 1.79 & 0.08 & 2.67 & 0.15 & 1.50 & 0.11 & 1.23 & 0.08 & 1.64 & 0.11 & 1.33 & 0.12 \\
Fe XXIV & 10.64 & 2.52 & 0.08 & 4.76 & 0.17 & 1.89 & 0.09 & 1.35 & 0.08 & 2.00 & 0.12 & 1.49 & 0.13 \\
Fe XXV & 1.85 & 0.43 & 0.12 & 2.02 & 0.35 & 4.69 & 1.58 & 0.22 & 0.12 & 0.27 & 0.20 & 1.22 & 1.12 \\
Fe XXVI & 1.78 & 0.05 & 0.17 & 0.67 & 0.46 & 12.94 & 44.59 & 0.00 & 0.19 & 0.06 & 0.32 & - & - \\
O VII & 21.60 & 4.25 & 1.68 & 0.02 & 2.88 & 0.00 & 0.68 & 1.48 & 1.79 & 0.82 & 2.60 & 0.55 & 1.88 \\
O VIII & 18.97 & 11.73 & 0.57 & 11.88 & 1.15 & 1.01 & 0.11 & 26.27 & 0.97 & 24.48 & 1.22 & 0.93 & 0.06 \\
Ne IX & 13.45 & 2.12 & 0.15 & 1.85 & 0.26 & 0.87 & 0.14 & 4.39 & 0.21 & 4.54 & 0.29 & 1.04 & 0.08 \\
Ne X & 12.13 & 7.70 & 0.20 & 8.56 & 0.39 & 1.11 & 0.06 & 16.04 & 0.33 & 15.86 & 0.43 & 0.99 & 0.03 \\
Mg XI & 9.17 & 0.79 & 0.06 & 0.92 & 0.10 & 1.16 & 0.15 & 0.74 & 0.06 & 0.86 & 0.09 & 1.18 & 0.15 \\
Mg XII & 8.42 & 1.82 & 0.07 & 2.88 & 0.14 & 1.58 & 0.10 & 1.46 & 0.08 & 1.64 & 0.12 & 1.12 & 0.10 \\
Al XII & 7.76 & 0.07 & 0.03 & 0.15 & 0.07 & 2.10 & 1.42 & 0.23 & 0.04 & 0.16 & 0.06 & 0.69 & 0.29 \\
Al XIII & 7.17 & 0.25 & 0.04 & 0.35 & 0.08 & 1.45 & 0.42 & 0.19 & 0.05 & 0.25 & 0.07 & 1.31 & 0.47 \\
Si XIII & 6.65 & 0.99 & 0.06 & 1.42 & 0.11 & 1.43 & 0.14 & 0.83 & 0.06 & 0.89 & 0.08 & 1.07 & 0.13 \\
Si XIV & 6.18 & 1.51 & 0.07 & 2.65 & 0.15 & 1.75 & 0.13 & 0.95 & 0.07 & 1.32 & 0.11 & 1.38 & 0.15 \\
S XV & 5.04 & 0.60 & 0.09 & 0.79 & 0.19 & 1.32 & 0.37 & 0.39 & 0.10 & 0.75 & 0.16 & 1.90 & 0.63 \\
S XVI & 4.73 & 0.52 & 0.09 & 0.87 & 0.20 & 1.66 & 0.48 & 0.53 & 0.11 & 0.66 & 0.17 & 1.25 & 0.41 \\
Ar XVII & 3.95 & 0.21 & 0.06 & 0.23 & 0.13 & 1.11 & 0.72 & 0.31 & 0.08 & 0.33 & 0.11 & 1.06 & 0.44 \\
Ar XVIII & 3.73 & 0.10 & 0.06 & 0.31 & 0.15 & 3.13 & 2.55 & 0.20 & 0.08 & 0.18 & 0.12 & 0.87 & 0.67 \\
Ca XIX & 3.18 & 0.21 & 0.06 & 0.38 & 0.12 & 1.84 & 0.77 & 0.19 & 0.07 & 0.24 & 0.10 & 1.21 & 0.68 \\
Ca XX & 3.02 & 0.04 & 0.06 & 0.28 & 0.16 & 6.83 & 11.42 & 0.20 & 0.09 & 0.00 & 0.11 & 0.00 & 0.53 \\
Ni XIX & 12.43 & 0.37 & 0.073 & 0.52 & 0.15 & 1.41 & 0.49 & 0.32 & 0.08 & 0.23 & 0.12 & 0.74 & 0.42 \\

\hline \multicolumn{14}{l}{{\it NOTE:} Fluxes of H-like ion lines
include both transitions of the
unresolved Ly-$\alpha$ doublet.}\\
\multicolumn{14}{l}{Fluxes of He-like ion lines include only the resonant transition.}\\
\multicolumn{14}{l}{Fluxes of L-shell Fe ion lines include all
transitions within $\pm$0.03~\AA\ of the specified wavelength.}

\end{tabular}
\end{table*}

\begin{table*}[!h]

\caption{\label{tab:line_fluxes3} Measured line fluxes in
10$^{-4}$ photons s$^{-1}$ cm$^{-2}$ of TZ CrB and CC Eri} 

\begin{tabular}{lc | cccccc | cccccc|}

\hline \hline
    &      &\multicolumn{6}{|c|}{{\bf TZ CrB}}                 &\multicolumn{6}{|c|}{{\bf CC Eri}}  \tabularnewline
    &      &\multicolumn{2}{|c}{Quiescence} & \multicolumn{2}{c}{Flare} & \multicolumn{2}{c|}{Flare/Quies.}
           &\multicolumn{2}{|c}{Quiescence} & \multicolumn{2}{c}{Flare} & \multicolumn{2}{c|}{Flare/Quies.} \tabularnewline
Ion & Wave & Flux   & Err & Flux & Err & Ratio & Err   & Flux & Err & Flux & Err & Ratio & Err   \\

\hline

Fe XVII & 15.01 & 12.76 & 0.18 & 14.20 & 0.83 & 1.11 & 0.07 & 2.49 & 0.09 & 7.72 & 0.75 & 3.09 & 0.32 \\
Fe XVIII & 14.21 & 5.04 & 0.12 & 6.91 & 0.69 & 1.37 & 0.14 & 0.86 & 0.06 & 3.43 & 0.64 & 3.97 & 0.80 \\
Fe XIX & 13.51 & 3.65 & 0.10 & 4.29 & 0.54 & 1.18 & 0.15 & 0.57 & 0.05 & 2.79 & 0.48 & 4.91 & 0.93 \\
Fe XX & 12.84 & 2.78 & 0.09 & 4.12 & 0.51 & 1.48 & 0.19 & 0.51 & 0.04 & 2.21 & 0.42 & 4.35 & 0.90 \\
Fe XXI & 12.28 & 2.47 & 0.08 & 3.49 & 0.44 & 1.41 & 0.18 & 0.41 & 0.03 & 2.03 & 0.40 & 4.95 & 1.06 \\
Fe XXII & 11.77 & 1.43 & 0.05 & 1.97 & 0.29 & 1.37 & 0.21 & 0.26 & 0.02 & 2.53 & 0.28 & 9.66 & 1.35 \\
Fe XXIII & 11.00 & 0.97 & 0.04 & 2.13 & 0.24 & 2.19 & 0.26 & 0.23 & 0.02 & 3.44 & 0.24 & 15.16 & 1.59 \\
Fe XXIV & 10.64 & 0.87 & 0.04 & 4.93 & 0.30 & 5.68 & 0.43 & 0.22 & 0.02 & 6.98 & 0.31 & 32.05 & 3.14 \\
Fe XXV & 1.85 & 0.06 & 0.04 & 1.68 & 0.57 & 27.42 & 20.26 & 0.03 & 0.03 & 3.15 & 0.59 & - & - \\
Fe XXVI & 1.78 & 0.01 & 0.07 & 0.17 & 0.84 & 12.13 & 81.06 & 0.01 & 0.05 & 0.67 & 0.77 & - & - \\
O VII & 21.60 & 2.48 & 0.35 & 0.71 & 6.56 & 0.29 & 2.65 & 3.66 & 0.41 & 7.28 & 7.85 & 1.99 & 2.16 \\
O VIII & 18.97 & 19.92 & 0.57 & 32.78 & 2.95 & 1.65 & 0.16 & 14.75 & 0.51 & 40.15 & 3.62 & 2.72 & 0.26 \\
Ne IX & 13.45 & 4.25 & 0.12 & 5.38 & 0.61 & 1.26 & 0.15 & 2.96 & 0.08 & 5.10 & 0.55 & 1.72 & 0.19 \\
Ne X & 12.13 & 9.28 & 0.15 & 15.54 & 0.78 & 1.67 & 0.09 & 4.96 & 0.10 & 16.81 & 0.74 & 3.39 & 0.16 \\
Mg XI & 9.17 & 1.54 & 0.04 & 1.70 & 0.22 & 1.10 & 0.15 & 0.27 & 0.01 & 1.01 & 0.15 & 3.71 & 0.58 \\
Mg XII & 8.42 & 1.67 & 0.04 & 4.07 & 0.25 & 2.43 & 0.16 & 0.28 & 0.02 & 3.37 & 0.25 & 12.12 & 1.21 \\
Al XII & 7.76 & 0.14 & 0.02 & 0.17 & 0.12 & 1.17 & 0.88 & 0.02 & 0.01 & 0.07 & 0.10 & 3.66 & 5.25 \\
Al XIII & 7.17 & 0.10 & 0.02 & 0.44 & 0.15 & 4.52 & 1.74 & 0.02 & 0.01 & 0.30 & 0.12 & 13.52 & 7.37 \\
Si XIII & 6.65 & 1.16 & 0.03 & 1.84 & 0.19 & 1.58 & 0.17 & 0.37 & 0.02 & 1.93 & 0.16 & 5.18 & 0.49 \\
Si XIV & 6.18 & 0.78 & 0.03 & 2.54 & 0.24 & 3.28 & 0.35 & 0.28 & 0.02 & 4.83 & 0.26 & 17.55 & 1.41 \\
S XV & 5.04 & 0.38 & 0.04 & 1.28 & 0.35 & 3.38 & 0.99 & 0.18 & 0.02 & 2.05 & 0.33 & 11.31 & 2.34 \\
S XVI & 4.73 & 0.14 & 0.04 & 1.07 & 0.37 & 7.79 & 3.39 & 0.10 & 0.02 & 2.53 & 0.37 & 24.91 & 6.53 \\
Ar XVII & 3.95 & 0.11 & 0.02 & 0.27 & 0.24 & 2.41 & 2.21 & 0.04 & 0.01 & 0.70 & 0.22 & 18.08 & 8.32 \\
Ar XVIII & 3.73 & 0.02 & 0.02 & 0.36 & 0.26 & 18.00 & 24.46 & 0.02 & 0.01 & 0.40 & 0.23 & 20.63 & 18.37 \\
Ca XIX & 3.18 & 0.07 & 0.02 & 0.40 & 0.22 & 5.83 & 3.59 & 0.00 & 0.01 & 0.33 & 0.20 & - & - \\
Ca XX & 3.02 & 0.01 & 0.02 & 0.35 & 0.27 & 25.25 & 42.87 & 0.01 & 0.01 & 0.22 & 0.22 & 20.97 & 33.24 \\
Ni XIX & 12.43 & 0.65 & 0.04 & 1.10 & 0.30 & 1.69 & 0.47 & 0.16 & 0.02 & 0.60 & 0.25 & 3.87 & 1.73 \\

\hline \multicolumn{14}{l}{{\it NOTE:} Fluxes of H-like ion lines
include both transitions of the
unresolved Ly-$\alpha$ doublet.}\\
\multicolumn{14}{l}{Fluxes of He-like ion lines include only the resonant transition.}\\
\multicolumn{14}{l}{Fluxes of L-shell Fe ion lines include all
transitions within $\pm$0.03~\AA\ of the specified wavelength.}

\end{tabular}
\end{table*}


\clearpage

\section{Measured Abundances}

\begin{table}[!ht]
\caption{\label{tab:II_Peg_abund} II Peg abundances relative to Fe
during flare and quiescence.}
\begin{tabular}{c c c c c c c}
\hline \hline
    & \multicolumn{2}{c}{Quies.} & \multicolumn{2}{c}{Flare} & \multicolumn{2}{c}{Ratio} \\
El. & X/Fe  &Error & X/Fe  & Error & Flare/Quies.& Error \\
\hline

O  & 133   & 10   & 97.2  & 9.6   & 0.73  & 0.09  \\
Ne & 51.56 & 2.5  & 44.5  & 1.6   & 0.864 & 0.053 \\
Mg & 2.92  & 0.26 & 2.75  & 0.19  & 0.94  & 0.10  \\
Al & 0.30  & 0.13 & 0.332 & 0.01  & 1.10  & 0.57  \\
Si & 2.39  & 0.22 & 2.015 & 0.15  & 0.84  & 0.10  \\
S  & 1.60  & 0.36 & 1.11  & 0.21  & 0.70  & 0.21  \\
Ar & 0.57  & 0.28 & 0.31  & 0.16  & 0.54  & 0.39  \\
Ca & N/A   & N/A  & 0.28  & 0.17  & N/A   & N/A   \\
Ni & 0.02  & 0.05 & 0.048 & 0.045 & 2.4   & 6     \\

\hline
\end{tabular}
\end{table}

\begin{table}[!ht]
\caption{\label{tab:OU_And_abund} OU And abundances relative to Fe
during flare and quiescence.}
\begin{tabular}{c c c c c c c}
\hline \hline
    & \multicolumn{2}{c}{Quies.} & \multicolumn{2}{c}{Flare} & \multicolumn{2}{c}{Ratio} \\
El. & X/Fe  &Error & X/Fe  & Error & Flare/Quies.& Error \\
\hline

O  & 22.4  & 4.3   & 16.1  & 3.0   & 0.72  & 0.19  \\
Ne & 7.47  & 0.62  & 4.00  & 0.36  & 0.536 & 0.065 \\
Mg & 1.98  & 0.16  & 1.80  & 0.13  & 0.906 & 0.097 \\
Al & 0.118 & 0.064 & 0.169 & 0.051 & 1.44  & 0.89  \\
Si & 1.59  & 0.12  & 1.244 & 0.086 & 0.781 & 0.079 \\
S  & 0.46  & 0.14  & 0.41  & 0.090 & 0.88  & 0.33  \\
Ar & 0.16  & 0.11  & 0.118 & 0.064 & 0.74  & 0.65  \\
Ca & 0.10  & 0.12  & 0.149 & 0.062 & 1.4   & 1.8   \\
Ni & N/A   & N/A   & 0.038 & 0.040 & N/A   & N/A   \\

\hline
\end{tabular}
\end{table}

\begin{table}[!ht]
\caption{\label{tab:Algol_abund} Algol abundances relative to Fe
during flare and quiescence.}
\begin{tabular}{c c c c c c c}
\hline \hline
    & \multicolumn{2}{c}{Quies.} & \multicolumn{2}{c}{Flare} & \multicolumn{2}{c}{Ratio} \\
El. & X/Fe  &Error & X/Fe  & Error & Flare/Quies.& Error \\
\hline

O  & 22.0  & 1.8   & 16.62 & 2.2 & 0.75    & 0.12 \\
Ne & 8.90  & 0.25  & 6.77  & 0.32 & 0.760  & 0.042 \\
Mg & 1.507 & 0.055 & 1.513 & 0.073 & 1.004 & 0.061 \\
Al & 0.168 & 0.027 & 0.167 & 0.036 & 0.99  & 0.27 \\
Si & 1.171 & 0.046 & 1.198 & 0.062 & 1.023 & 0.066 \\
S  & 0.500 & 0.06  & 0.400 & 0.068 & 0.80  & 0.16 \\
Ar & 0.172 & 0.048 & 0.133 & 0.050 & 0.77  & 0.36 \\
Ca & 0.215 & 0.06  & 0.184 & 0.054 & 0.85  & 0.34 \\
Ni & 0.063 & 0.013 & 0.080 & 0.023 & 1.27  & 0.45 \\

\hline
\end{tabular}
\end{table}

\begin{table}[!ht]
\caption{\label{tab:HR1099_abund} HR 1099 abundances relative to
Fe during flare and quiescence.}
\begin{tabular}{c c c c c c c}
\hline \hline
    & \multicolumn{2}{c}{Quies.} & \multicolumn{2}{c}{Flare} & \multicolumn{2}{c}{Ratio} \\
El. & X/Fe  &Error & X/Fe  & Error & Flare/Quies.& Error \\
\hline

O  & 74.1  & 3.4   & 51.0 & 3.1  & 0.69 & 0.05 \\
Ne & 25.23 & 0.67  & 19.0 & 0.6  & 0.75 & 0.03 \\
Mg & 1.77  & 0.09  & 1.58 & 0.10 & 0.89 & 0.07 \\
Al & 0.29  & 0.05  & 0.24 & 0.05 & 0.82 & 0.22 \\
Si & 1.26  & 0.07  & 1.20 & 0.08 & 0.95 & 0.08 \\
S  & 0.65  & 0.11  & 0.75 & 0.12 & 1.15 & 0.27 \\
Ar & 0.48  & 0.10  & 0.35 & 0.11 & 0.72 & 0.27 \\
Ca & 0.44  & 0.14  & 0.31 & 0.14 & 0.68 & 0.38 \\
Ni & 0.065 & 0.017 & 0.04 & 0.02 & 0.61 & 0.35 \\

\hline
\end{tabular}
\end{table}

\begin{table}[!ht]
\caption{\label{tab:TZ_CrB_abund} TZ CrB abundances relative to Fe
during flare and quiescence.}
\begin{tabular}{c c c c c c c}
\hline \hline
    & \multicolumn{2}{c}{Quies.} & \multicolumn{2}{c}{Flare} & \multicolumn{2}{c}{Ratio} \\
El. & X/Fe  &Error & X/Fe  & Error & Flare/Quies.& Error \\
\hline

O  & 25.1  & 1.7   & 26.0  & 3.6   & 1.04  & 0.16 \\
Ne & 7.95  & 0.38  & 8.15  & 0.44  & 1.025 & 0.073 \\
Mg & 1.66  & 0.07  & 1.71  & 0.10  & 1.026 & 0.076 \\
Al & 0.138 & 0.015 & 0.172 & 0.054 & 1.25  & 0.41 \\
Si & 1.25  & 0.06  & 1.176 & 0.090 & 0.943 & 0.085 \\
S  & 0.492 & 0.055 & 0.53  & 0.12  & 1.08  & 0.28 \\
Ar & 0.240 & 0.055 & 0.142 & 0.080 & 0.59  & 0.36 \\
Ca & 0.29  & 0.09  & 0.177 & 0.080 & 0.60  & 0.33 \\
Ni & 0.057 & 0.005 & 0.076 & 0.021 & 1.33  & 0.38  \\

\hline
\end{tabular}
\end{table}

\begin{table}[!ht]
\caption{\label{tab:CC_Eri_abund} CC Eri abundances relative to Fe
during flare and quiescence.}
\begin{tabular}{c c c c c c c}
\hline \hline
    & \multicolumn{2}{c}{Quies.} & \multicolumn{2}{c}{Flare} & \multicolumn{2}{c}{Ratio} \\
El. & X/Fe  &Error & X/Fe  & Error & Flare/Quies.& Error \\
\hline

O  & 72.3  & 3.3   & 34.6  & 4.1   & 0.48  & 0.06  \\
Ne & 21.2  & 0.5   & 9.8   & 0.5   & 0.465 & 0.025 \\
Mg & 1.49  & 0.07  & 1.34  & 0.10  & 0.90  & 0.08  \\
Al & 0.13  & 0.04  & 0.10  & 0.04  & 0.78  & 0.37  \\
Si & 2.17  & 0.09  & 1.49  & 0.08  & 0.689 & 0.046 \\
S  & 1.26  & 0.15  & 0.76  & 0.09  & 0.60  & 0.10  \\
Ar & 0.40  & 0.12  & 0.17  & 0.05  & 0.44  & 0.19  \\
Ca & 0.08  & 0.14  & 0.10  & 0.05  & 1.2   & 2.1   \\
Ni & 0.074 & 0.012 & 0.077 & 0.033 & 1.05  & 0.48  \\

\hline
\end{tabular}
\end{table}

\end{appendix}

\end{document}